\documentclass{iopart}
\usepackage{iopams}
\expandafter\let\csname equation*\endcsname\relax
\expandafter\let\csname endequation*\endcsname\relax
\usepackage{titlesec}
\titleformat{\chapter}[hang] 
{\normalfont\huge\bfseries}{\thechapter}{1em}{} 

\usepackage[hidelinks]{hyperref}
\usepackage{amsmath}
\usepackage{amsthm}
\usepackage{amssymb}
\usepackage{amsfonts}
\usepackage{dsfont}
\usepackage[numbers,square,sort&compress]{natbib}

\usepackage{graphicx}
\newcommand{\nn}{{\nonumber}}

\begin{document}
\title[An Interfering Mean-Field Propagator for the Bose-Hubbard Dimer]{Action on the Sphere: An Interfering Mean-Field Propagator for the Bose-Hubbard Dimer}
\author{Elana F. Todd-Miller and Eva-Maria Graefe$^*$} 

\address{Department of Mathematics, Imperial College London, London SW7 2AZ,
United Kingdom}
\address{\small $^*$ Author to whom any correspondence should be addressed}
\ead{e.m.graefe@imperial.ac.uk}

\begin{abstract}
    The Bose-Hubbard system has been studied extensively both theoretically and experimentally, in particular in the context of ultracold atomic gases in optical lattices. Even in the two-mode case the many-particle dynamics display complex interference effects resulting in collapse and revival phenomena as well as tunnelling. The most basic theoretical description is the mean-field approximation, which can be derived from a time-dependent variational principle assuming the many-particle wave function is an $\mathrm{SU}(2)$ coherent state. Here we build on this to construct a simple initial-value coherent state propagator, summing over mean-field trajectories and keeping track of their phases, given by the corresponding mean-field actions. This yields an approximation to the full time-dependent many-particle state, and is able to reproduce collapse and revival dynamics. Applying a time-slicing procedure on top of this, we are able to accurately capture many-particle tunnelling effects. While in this paper we focus our analysis on the Bose-Hubbard dimer, the  methods developed can be applied to more general $\mathrm{SU}(2)$ Hamiltonians, and can be extended to $\mathrm{SU}(M)$ systems.
\end{abstract}
\section{Introduction}

The correspondence between many-particle and mean-field descriptions of multi-boson quantum systems is formally identical to that of the quantum and classical descriptions of a single particle. This fact has led to a resurgence of semiclassical methods adapted and applied to many-particle quantum systems. One of the most prominent examples is the Bose-Hubbard model, which can be experimentally realised using ultracold bosons in optical lattices, and  displays a plethora of non-trivial many-body effects \cite{fisher1989boson,greiner2002collapse,greiner2002quantum,bloch2005ultracold, jaksch2005cold,polkovnikov2011colloquium, engl2014coherent, kolovsky2016bose,tomsovic2018post,richter2022semiclassical}. These systems are promising candidates for practical applications, as quantum simulators \cite{gross2017quantum,daley2022practical}. 
Already the two-mode Bose-Hubbard Hamiltonian, modelling for example an ensemble of cold atoms in a double-well potential \cite{eilbeck1985discrete, scott1986quantized, milburn1997quantum,albiez2005direct,gati2007bosonic,zibold2010classical},  displays characteristic collapse and revival phenomena \cite{milburn1997quantum,schlagheck2022resurgent} and many-particle tunnelling between mean-field self-trapping states \cite{kalosakas2003multiple,pudlik2014tunneling}. Though such effects are different from those which appear in higher mode Bose-Hubbard models, such as the superfluid-to-insulator transition, they are an ideal platform for the development of semiclassical techniques.

The mean-field approximation of the full many-particle dynamics, is often derived using multi-mode Glauber coherent states, yet, an $N$-boson condensed state is mathematically correctly described by an $\mathrm{SU}(M)$ coherent state \cite{radcliffe1971some,perelomov1977generalized, zhang1990coherent}. In the limit of large particle numbers, the two coherent-state approximations yield equivalent mean-field dynamics, but only the treatment via $\mathrm{SU}(M)$ coherent states correctly preserves the $U(1)$ symmetry of the many-particle system, leading to mean-field dynamics on a compact phase space, reflecting the well-defined particle number of the many-particle system \cite{buonsante2008some,trimborn2008exact,trimborn2009beyond,qiao2024quench}. While the mean-field dynamics is often a sufficient approximation for the dynamical behaviour of cold atoms in optical lattices for large numbers of particles and high filling factors, strong interactions and smaller particle numbers can lead to a breakdown of the mean-field approximation \cite{vardi2001bose,esteve2008squeezing,muessel2014scalable, engl2014coherent,tomsovic2018post}. Taking the initial quantum fluctuations into account leads to the commonly applied truncated Wigner approximation \cite{steel1998dynamical,sinatra2002truncated,polkovnikov2010phase,polkovnikov2011colloquium}, which essentially propagates an ensemble of initial mean-field trajectories and sums them up incoherently, i.e., excluding phase factors. This can reproduce the typical collapse observed in the expectation values in the presence of interaction. Neither the subsequent revival behaviour nor many-particle tunnelling effects can be captured by this approximation. 

To go beyond the truncated Wigner approximation one needs to take the phases of the individual mean-field trajectories into account. In this spirit, the Herman-Kluk semiclassical propagator, based on the flat space multi-mode Glauber coherent states, has been shown to accurately capture collapse and revival dynamics in the single \cite{ray2016dynamics}, and two- and three-mode \cite{simon2014time} Bose-Hubbard model. Further \cite{engl2014coherent,engl2016semiclassical,tomsovic2018post,mathew2019semiclassical} use the Glauber coherent state basis to derive full Van Vleck type saddle-point approximations to a path integral propagator. Although this is a powerful approach, its implementation is highly non-trivial, and the recovery of the full quantum state beyond single mode systems would require major effort. 

A tangential approach to the semiclassical treatment, uses a time-dependent variational principle with a linear combination of coherent states. A single-coherent state approximation yields the mean-field dynamics \cite{wimberger2021finite}; a coupled coherent state approximation  \cite{shalashilin2001description, shalashilin2008gaussian} can be employed to simulate many-body dynamics to arbitrary accuracy \cite{qiao2021exact,qiao2023revealing}. For this purpose a time-dependent variational principle is applied to a wavefunction expressed as a superposition of $\mathrm{SU}(2)$ coherent states, leading to coupled equations of motion for the coefficients of each coherent state in addition to the mean-field dynamics of the individual states. In principle this can exactly reproduce the full many-body state, however, the computational cost grows significantly as more coherent states are added to the ansatz.

Here we follow a different route, closer in spirit to semiclassical ideas, propagating a set of uncoupled $\mathrm{SU}(2)$ coherent states, similar to the Gaussian propagator for flat phase spaces \cite{heller1981frozen,littlejohn1986semiclassical}. For this purpose, we apply the time evolution operator to an arbitrary state expressed in the overcomplete basis of coherent states. The evolution of each component state is then approximated using the mean-field equations of motion, crucially keeping track of each one's phase which is given by the action of the corresponding mean-field trajectory. This yields an approximation of the full many-particle state using only mean-field dynamics, in the spirit of an $\mathrm{SU}(M)$ initial value coherent state propagator \cite{viscondi2011initial}. Numerically we implement the integral on the sphere using Gauss-Legendre quadrature. Alternatively the integral could be handled with a Monte Carlo approach. The resulting propagator is exact in the non-interacting case, and is highly effective in capturing the collapse and revival phenomena displayed in the full many-body dynamics. While the grid implementation may not be feasible for systems with larger phase-space dimension, the method does not involve coupled equations of motion between the different component states, and is thus highly parallelisable. 

This propagator is extended using a time-slicing procedure, similar to the well-known procedure in the flat case \cite{huber1988hybrid, bergold2022error}. This method is computationally faster, and most importantly is able to replicate even many-particle tunnelling dynamics. We note that in the time-slicing, we pick up an accumulated effect of an infinitesimal error in the propagator, which we are able to estimate from analysis of the pure interaction term in the semiclassical limit. Interestingly, this manifests as a scaling factor in the interaction term, the asymptotic behaviour of which we derive analytically, and which we numerically find for smaller values of the particle number.  Incorporating this scaling factor, the states propagated with the time-sliced interfering mean-field propagator are almost indistinguishable from the true dynamics even for long propagation times. 

The paper is organised as follows. We begin in section \ref{sec-BH-SU(2)} with a short introduction to the Bose-Hubbard dimer and its relation to the $\mathrm{su}(2)$ algebra, and provide the necessary background on $\mathrm{SU}(2)$ coherent states. We briefly review how the mean-field dynamics is obtained from a coherent-state approximation via a time-dependent variational principle, including a phase factor. In section \ref{sec-MFP} we introduce the Interfering Mean-Field (IMF) propagator. For the pure interaction Hamiltonian, a closed form expression for the matrix elements of the propagator in the $\hat J_z$ basis can be found. Using a saddle-point approximation we find a reduced expression in the limit of large particle number $N$, which differs from the exact result. Nevertheless, as we show through examples, this yields a very good approximation for expectation values and even the full time-evolved Wigner functions. We discuss the numerical implementation of this propagator for a general Hamiltonian and show that it reproduces break-down and revival behaviour, but stops short of tunnelling. 
We then introduce a time-sliced version of the IMF propagator in section \ref{sec-TS}. We show that this can reproduce tunnelling behaviour to high accuracy, even far from the semiclassical limit, with a scaling factor in the interaction parameter. Based on the results of the saddle-point approximation from the previous section we predict this factor in the limit of large $N$. We further use numerical optimisation to find the factors for a range of small $N$. 
We finish with a summary and outlook in section \ref{sec-SO}. Some technical details are left for the appendices.

\section{\texorpdfstring{Bose-Hubbard dimer and $\mathrm{SU}(2)$ coherent states}{Bose-Hubbard dimer and SU(2) coherent states}}
\label{sec-BH-SU(2)}

A two-mode bosonic system with on-site two-body interaction may be described by a Bose-Hubbard Hamiltonian of the form
\begin{equation}
    \hat{H}_{BH} = \varepsilon(\hat{n}_1 - \hat{n}_2) + \nu (\hat{a}_1^\dagger\hat{a}_2 + \hat{a}_1\hat{a}_2^\dagger) + c\left(\hat{n}_1(\hat n_1-1)+\hat{n}_2(\hat n_2-1)\right), 
    \label{BHHamfull}
\end{equation}
where $\hat{a}_k, \hat{a}_k^\dagger$ are the bosonic annihilation and creation operators for the $k$-th mode, and $\hat{n}_k = \hat{a}_k^\dagger\hat{a}_k$ are the number operators associated to each site. The coefficient $\epsilon$ denotes the energy difference between the two modes, $\nu$, the single particle hopping amplitude, and $c$ the on-site interaction strength. The Hamiltonian commutes with the overall particle number $\hat N=\hat n_1+\hat n_2$, which is thus dynamically conserved. 
For a fixed particle number $N=n_1+n_2$, the Hilbert space is $(N+1)$-dimensional with basis Fock states $|n_1, n_2\rangle$, with $\hat n_k|n_1,n_2\rangle=n_k|n_1,n_2\rangle$.

Near zero temperature, all bosons populate the same single-particle state, forming a condensate. This condensed state can be expressed as
\begin{equation}
|\psi_1,\psi_2\rangle=\frac{1}{\sqrt{N!}}\left(\psi_1\hat a_1^\dagger+\psi_2\hat a_2^\dagger\right)^N|0,0\rangle,
\label{consend}
\end{equation}
where $\psi_1,\psi_2\in\mathds{C}$ are the amplitudes of the two-mode single particle state.

The system can equivalently be described as a spin $j=N/2$ system via the Jordan-Schwinger transformation
\cite{zhang1990coherent}
\begin{align}
    \hat{J}_x &= \frac{1}{2}(\hat{a}_1^\dagger \hat{a}_2 +\hat{a}_1 \hat{a}_2^\dagger), \\
    \hat{J}_y &= \frac{1}{2\rmi}(\hat{a}_1^\dagger \hat{a}_2 - \hat{a}_1 \hat{a}_2^\dagger), \\
    \hat{J}_z &= \frac{1}{2}(\hat{n}_1 -\hat{n}_2).
\end{align}
With these, the raising and lowering operators may be defined as $\hat{J}_\pm = \hat{J}_x \pm \rmi \hat{J}_y.$ These operators satisfy the standard $\mathrm{su}(2)$ commutation relations
\begin{equation}
[\hat J_k,\hat J_l]=\rmi\epsilon_{klm}\hat J_m, \quad [\hat{J}_+, \hat{J}_-] = 2 \hat{J}_z, \quad [\hat{J}_z, \hat{J}_\pm] = \pm \hat{J}_\pm.
\end{equation}
Note that we will use rescaled units with $\hbar=1$ throughout this paper. The inverse particle number, i.e. the inverse spin quantum number will act as effective $\hbar$.

In terms of the spin operators, up to overall terms proportional to the fixed particle number $\hat N$, the Bose-Hubbard Hamiltonian (\ref{BHHamfull}) becomes
\begin{equation}
    \hat{H}_{BH} = 2\varepsilon\hat{J}_z + 2\nu \hat{J}_x + 2\frac{\kappa}{j}\hat{J}_z^2,
    \label{BHHamdimer}
\end{equation}
where we have introduced $\kappa=N c$, such that $\kappa$ remains fixed in the macroscopic limit of $N \to \infty$. This ensures that the interaction energy per particle stays constant while the system size grows, permitting the comparison of different spin systems as well as resulting in well-defined classical dynamics. In this formulation the Fock states $|n_1, n_2 \rangle$ can be identified with the $\hat{J}_z$ eigenbasis, $|j, m\rangle$ where $j=N/2$ and $m=(n_1-n_2)/2$, and the condensed states are none other than the $\mathrm{SU}(2)$ or spin coherent states.

\subsection{\texorpdfstring{$\mathrm{SU}(2)$ coherent states}{SU(2) coherent states}}
$\mathrm{SU}(2)$ coherent states are commonly defined as \cite{radcliffe1971some, zhang1990coherent,perelomov1977generalized}
\begin{align}
    |\zeta\rangle := |\zeta;j\rangle = \frac{1}{(1+|\zeta|^2)^j}e^{\zeta\hat{J}_-} |j, j\rangle, 
\end{align}
where $\zeta  = e^{\rmi \phi}\tan\frac{\theta}{2}$ is a stereographic projection of the unit sphere with spherical co-ordinates $(\theta,\phi)$, onto the flat complex plane. Alternatively, the coherent states may be expressed in terms of the canonical conjugate variables  $z = \cos\theta$, and $\phi$. In the $\hat J_z$ basis, coherent states have the form
\begin{align}
\nn |\zeta\rangle &=(1+|\zeta|^2)^{-j}\sum_{m=-j}^{j}\sqrt{\binom{2j}{j-m}}\zeta^{j-m}|j,m\rangle, \\
&=\frac{1}{2^j}\sum_{m=-j}^{j} \sqrt{\binom{2j}{j-m}}\,(1+z)^{(j+m)/2} (1-z)^{(j-m)/2}  e^{\rmi \phi(j-m)}|j,m \rangle. \label{thisone}
\end{align}
A short calculation (given in appendix \ref{firstapp}) shows that this is equivalent to a condensed state (\ref{consend}) with
\begin{equation}
    \psi_1=\sqrt{\frac{1+z}{2}}, \quad \psi_2=\sqrt{\frac{1-z}{2}}\rme^{\rmi \phi}.
\end{equation}
The $\mathrm{SU}(2)$ coherent states form an overcomplete basis on the $(2j+1)$‑dimensional Hilbert space, forming the resolution of the identity \cite{zhang1990coherent}
\begin{align}
    \nn\hat{I} &= \frac{2j+1}{\pi} \int_\mathds{c} \frac{|\zeta \rangle \langle \zeta |}{(1+|\zeta|^2)^2} \rmd^2\zeta\\
&= \frac{2j+1}{4\pi} \int^{2\pi}_{0} \int^{1}_{-1}|z, \phi \rangle \langle z, \phi |  \rmd z \rmd\phi.
    \label{resofidentity_zphi}
\end{align}

The expectation values of the spin operators in $\mathrm{SU}(2)$ coherent states are given by
\begin{align}
    \langle \hat{J}_x \rangle &= \frac{j(\zeta + \zeta^*)}{1+|\zeta|^2} = j\sqrt{1 - z^2}\ \cos\phi, \\
    \langle \hat{J}_y \rangle &= \frac{j(\zeta - \zeta^*)}{\rmi(1+|\zeta|^2)} = j \sqrt{1 - z^2}\, \sin\phi , \\
    \langle \hat{J}_z \rangle &= j\frac{ 1 - |\zeta|^2 }{1 + | \zeta|^2}=j z,
    \label{Jminusexp}
\end{align}
We will frequently use scaled expectation values 
\begin{equation}
    x = \frac{\langle \hat{J}_x\rangle}{j}, \quad y = \frac{\langle \hat{J}_y\rangle}{j}, \quad z = \frac{\langle \hat{J}_z\rangle}{j}.
\end{equation} 
Expectation values of products of spin operators can be deduced from the group theoretical properties of the coherent states (see \ref{sec-CS-expval}).

\subsection{Time-dependent variational principle for a single SU(2) coherent state}
The assumption that the state remains condensed for all times, and can be described by a single particle mean-field state, is equivalent to a coherent-state approximation. The corresponding equations of motion can be derived from a time-dependent variational principle. 

Starting from the time-dependent Schrödinger equation
\begin{align}
     \rmi \frac{\partial }{\partial t}| \psi(t) \rangle &= \hat{H} |\psi(t) \rangle,
\end{align}
for an initially coherent state $|\psi_0\rangle = c_0 | \zeta_0\rangle$, 
constraining the state to remain coherent over time up to a phase factor,
\begin{align}
    | \psi(t) \rangle = c(t) | \zeta(t)\rangle \quad\text{ such that }\quad | \psi(0) \rangle = c_0 | \zeta_0\rangle, \label{constraint}
\end{align}
yields equations of motion for the coherent state coordinate $\zeta(t)$ and the prefactor $c(t)$, given by 
\begin{align}
    \rmi  \dot{\zeta} &= \frac{(1 + |\zeta|^2)^2}{2j} \frac{\partial \langle \hat{H}\rangle}{\partial \zeta^*}, \label{muEOM}\\
    \rmi  \dot{c} &= c \Big[ \langle \hat{H}\rangle + \rmi  \frac{j}{1 + |\zeta|^2}( \zeta^* \dot{\zeta} - \zeta \dot{\zeta}^*)\Big] \label{cEOMwithzeta}
\end{align} 
where we have used the abbreviation $\langle \hat H\rangle=\langle \zeta|\hat H|\zeta\rangle$ to denote the expectation value of the Hamiltonian in coherent states.
We will dub these the single coherent state time dependent variational principle (SCS-TDVP) equations of motion.
For details of the derivation see  \ref{sec-appendix-TDVP}.

Alternatively, we may write the equation for $\dot{c}$ as 
\begin{align}
    \rmi  \dot{c} &= c \Big[ \langle \hat{H}\rangle - \frac{1 + |\zeta|^2}{2}( \zeta \frac{\partial \langle \hat{H}\rangle}{\partial \zeta } + \zeta^* \frac{\partial \langle \hat{H}\rangle}{\partial \zeta^*})\Big] .\label{cEOMwithderiv}
\end{align} 
In terms of the real coordinates $z = \cos\theta $ and  $\phi$ with $\zeta =\tan\left(\frac{\theta}{2}\right)\rme^{\rmi\phi}$, the SCS-TDVP equations of motion take the form
\begin{align}
    \dot{\phi} &=\frac{1}{j}\frac{\partial \langle \hat{H}\rangle}{\partial z},\label{phiEOM}\\
    \dot{z} &= -\frac{1}{j}\frac{\partial \langle \hat{H}\rangle}{\partial \phi}, \label{zEOM}\\
    \rmi  \dot{c} &= c\Big[\langle \hat{H}\rangle + (1-z)\frac{\partial \langle \hat{H}\rangle}{\partial z}\Big]. \label{cEOM}
\end{align}
That is, the variables $(\phi,z)$ follow Hamiltonian dynamics where $\frac{1}{ j}\langle \hat{H} \rangle $ acts as Hamiltonian function. Expressing $c$ as 
\begin{equation}
    c(t) = c_0 e^{\rmi S(t; \zeta_0)}
\end{equation}
we recognise $S$ as the classical action, which can be expressed as
\begin{align}
    S(t;\zeta_0) &= \int^{t}_0\rmi \frac{ j}{1 + |\zeta(t')|^2}(\zeta^*(t')\dot{\zeta}(t') - \zeta(t')\dot{\zeta}^*(r')) - \langle \hat{H} \rangle(t')\rmd t', \\
    &= \int^t_0   j \dot{\phi}(t')[z(t') - 1] - \langle \hat{H} \rangle(t')\rmd t'.
\end{align}
For a Hamiltonian that is linear in the $\mathrm{SU}(2)$ generators $\hat J_k$, an initial coherent state remains coherent for all times, and thus the SCS-TDVP equations of motion (\ref{phiEOM})-(\ref{cEOM}) recover the exact dynamics. For more general Hamiltonians, such as the Bose-Hubbard dimer (\ref{BHHamdimer}) the time-dependent state (\ref{constraint}) is an approximation of the true quantum state, closely related to the mean-field approximation. 

Note that the equations of motion for $z$ and $\phi$ have been derived for the specific case of a Bose-Hubbard dimer using an equivalent approach in \cite{wimberger2021finite}. The single coherent state ansatz here is a special case of the coupled coherent state approach of \cite{qiao2021exact, qiao2023revealing}. 

Let us now derive the SCS-TDVP equations of motion for the  Bose-Hubbard Hamiltonian (\ref{BHHamdimer}). Using that in coherent states it holds that
\begin{align}
    \langle \hat{J}_k^2 \rangle &= \langle \hat{J}_k\rangle^2(1 - \frac{1}{2j}) + \frac{j}{2}.
\end{align}
one finds
\begin{equation}
        \frac{\langle \hat H_{BH} \rangle}{j }  = 2\varepsilon z + 2 \nu \cos \phi (1 - z^2)^{1/2} + 2 \kappa(1 - \frac{1}{2j}) z^2,
\end{equation}
where we have omitted an additive constant $\kappa$, which does not affect the dynamics and only adds an overall constant phase, independent of the initial conditions. 
The constraint-coherent state dynamics are governed by the equations of motion 
\begin{align}
     \dot{\phi} &= 2 \varepsilon  -\frac{2 \nu z \cos \phi }{(1 - z^2)^{1/2}} + 4 \kappa (1 - \frac{1}{2j}) z, \label{BHEOMphi}\\
     \dot{z} &= 2 \nu (1 - z^2)^{1/2} \sin \phi, \label{BHEOMz} \\
    \dot{c} &= - \rmi j c \Big[ 2 \varepsilon +\frac{2 \nu (1 - z^2)^{1/2}\cos \phi}{1 + z} -  2 \kappa(1 - \frac{1}{2j}) z(z - 2)\Big]. \label{BHEOMc}
\end{align}
As expected, in the limit of $j \to \infty$ the equations for $\phi$ and $z$ reduce to the well-known mean-field dynamics \cite{milburn1997quantum}, which, in the linear case $\kappa=0$ describe the exact dynamics for an initial coherent state.

\section{\texorpdfstring{The Interfering Mean-Field Propagator using $\mathrm{SU}(2)$ coherent states}{The Interfering Mean-Field Propagator using SU(2) coherent states}}
\label{sec-MFP}

Using the overcompleteness of $\mathrm{SU}(2)$ coherent states (\ref{resofidentity_zphi}), we can build on the SCS-TDVP to construct a simple ``semiclassical'' propagator in the spirit of Heller's frozen Gaussian propagator \cite{heller1981frozen}. For linear Hamiltonians, this yields the exact many-particle time evolution for arbitrary initial states, and for nonlinear (i.e. interacting) Hamiltonians yields a good approximation of the exact dynamics; in particular it is able to replicate coherent collapse and revival phenomena.

An arbitrary time evolved state $|\psi(t)\rangle = \hat{U}(t)|\psi(0)\rangle$ may be expanded as
\begin{equation}
    \hat{U}(t)|\psi(0)\rangle = \frac{2j+1}{4 \pi}\int^{2\pi}_{0}\int^{1}_{-1}  \hat{U}(t) |z_0, \phi_0 \rangle \langle z_0, \phi_0 |\psi(0)\rangle  \rmd z_0  \rmd\phi_0.
    \label{U_CS_int}
\end{equation}
This is an exact representation of the quantum dynamics. This can now be approximated by using the SCS-TDVP dynamics for the time-evolution of each initial coherent state $|z_0,\phi_0\rangle$ in the integral. That is, we replace 
$\hat U(t)|z_0,\phi_0\rangle$ with the SCS-TDVP approximation. However, instead of using the coherent state expectation value $\langle \hat H\rangle$ of the quantum Hamiltonian as generator in the equations of motion, we use the \textit{classical/mean-field limit} of the Hamiltonian, given by 
\begin{equation}
    H = \lim_{j\to\infty} \frac{\langle \hat{H}\rangle }{j}.
\end{equation}
This omission of the higher order $j$ corrections in the SCS-TDVP is to avoid overcounting of uncertainty effects that are described by these terms, but also encoded into the dynamics of neighbouring initial states in the integral (\ref{U_CS_int}). We have verified numerically  that using the SCS-TDVP trajectories with the correction term leads to consistently worse correspondence with the numerically exact many-particle dynamics.

The Interfering Mean-Field (IMF) propagator is thus defined as
\begin{equation}
    \hat{U}^{IMF}(t) := \frac{2j+1}{4\pi}\int^{2\pi}_{0}\int^{1}_{-1}  c(t;z_0, \phi_0)|z(t; z_0, \phi_0),\phi(t;z_0,\phi_0)\rangle \langle z_0, \phi_0 |  \rmd z_0  \rmd\phi_0,
    \label{U_IM_general}
\end{equation}
where $z(t)$, $\phi(t)$, and $c(t)$ are the solutions to the equations of motion 
\begin{equation}
    \dot{\phi} =\frac{\partial H}{\partial z}, \qquad
    \dot{z} = -\frac{\partial H}{\partial \phi}, \qquad
    \rmi  \dot{c} = jc\Big[H + (1-z)\frac{\partial H}{\partial z}\Big]. \label{allEOM_IM}
\end{equation}
at time $t$, with initial conditions $z(t=0)=z_0$, $\phi(t=0)=\phi_0$, and $c(t=0)=1$. 

For the Bose-Hubbard Hamiltonian considered here we have (up to an additive constant)
\begin{equation}
H(z,\phi)= 2\varepsilon z + 2 \nu \cos \phi (1 - z^2)^{1/2} + 2 \kappa z^2,
\end{equation}
the equations of motion are given by the mean-field equations 
\begin{align}
     \dot{\phi} &= 2 \varepsilon  -\frac{2 \nu z \cos \phi }{(1 - z^2)^{1/2}} + 4 \kappa z, \label{BHEOMphi_limit}\\
     \dot{z} &= 2 \nu (1 - z^2)^{1/2} \sin \phi, \label{BHEOMz_limit} \\
    \dot{c} &= - \rmi j  \Big[ 2 \varepsilon  +\frac{2 \nu (1 - z^2)^{1/2}\cos \phi}{1 + z} -  2 \kappa z(z - 2)\Big]c.\label{BHEOMc_limit}
\end{align}
In what follows, we consider initially coherent states for our examples as they yield interesting and clear dynamical behaviour. The method is in no way restricted to such states and performs equally well on arbitrary initial states.

\subsection{Pure Interaction Case}

\begin{figure}[tb]
  \centering
  \includegraphics[width=0.35\textwidth]{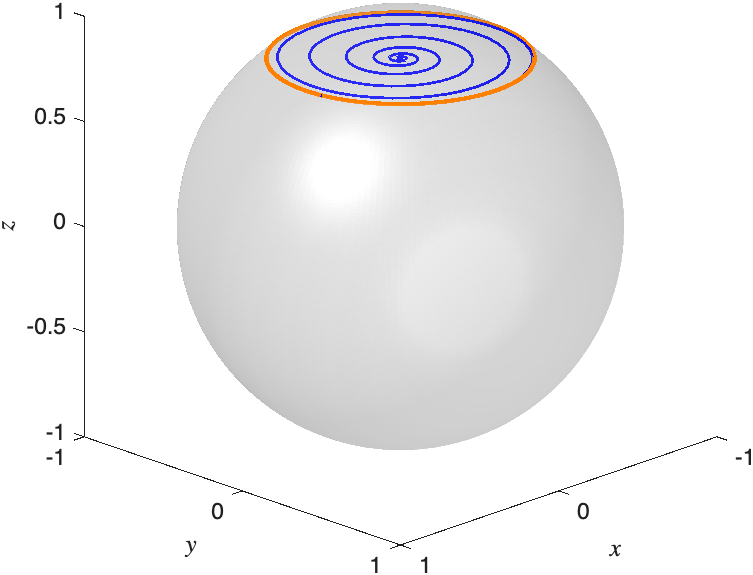}
  \hspace{0.3em}
  \includegraphics[width=0.55\textwidth]{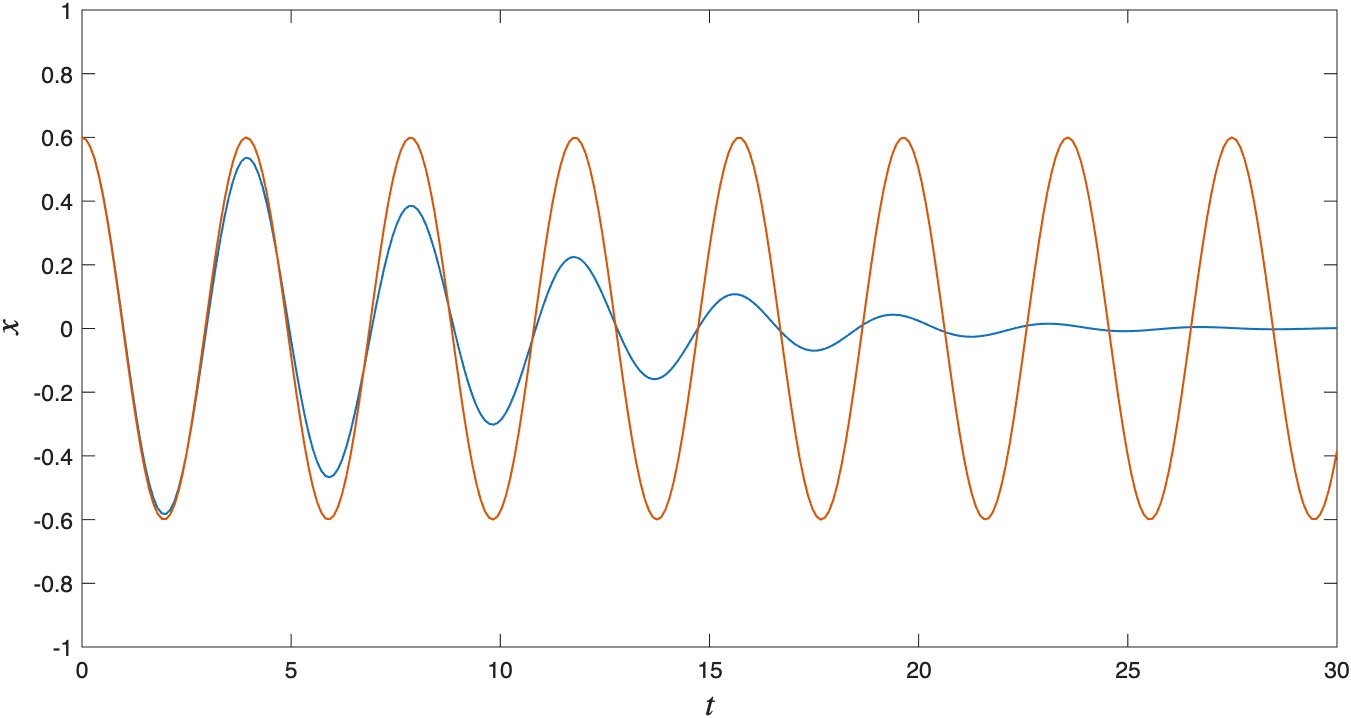}
  \caption{Quantum angular momentum expectation values (blue) and mean-field trajectory (orange) of a spin 50 system with $\kappa=0.5$, for an initial coherent state centred at  $z_0=0.8$ and $\phi_0=0$. The expectation values on the Bloch sphere are shown on the left and the x-expectation values as a function of time  on the right. }
  \label{fig:kerrclassic}
\end{figure}

\begin{figure}[tb]
  \centering
    \centering
    \includegraphics[width=0.65\textwidth]{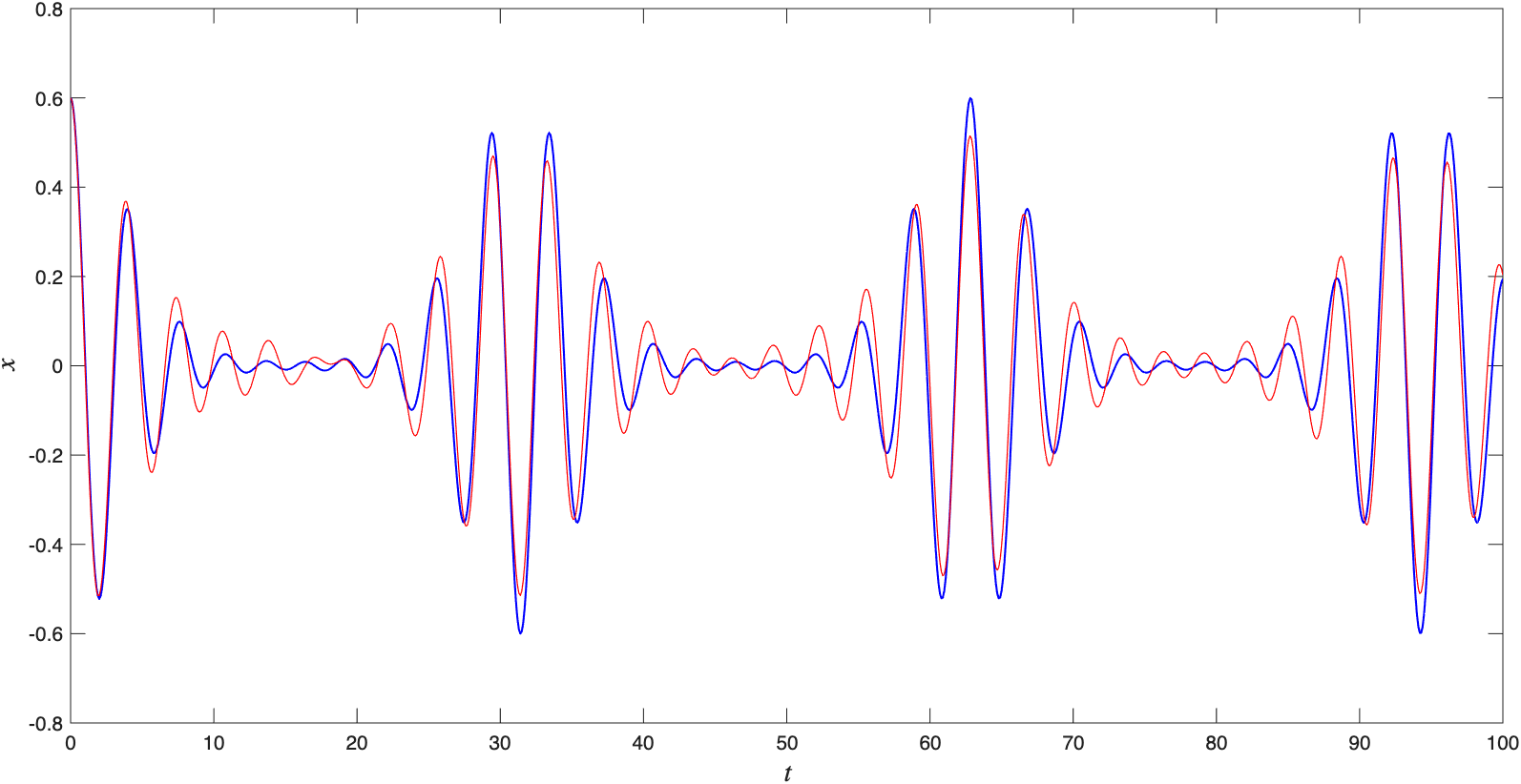}
    \vspace{0.3em}
    \includegraphics[width=0.65\textwidth]{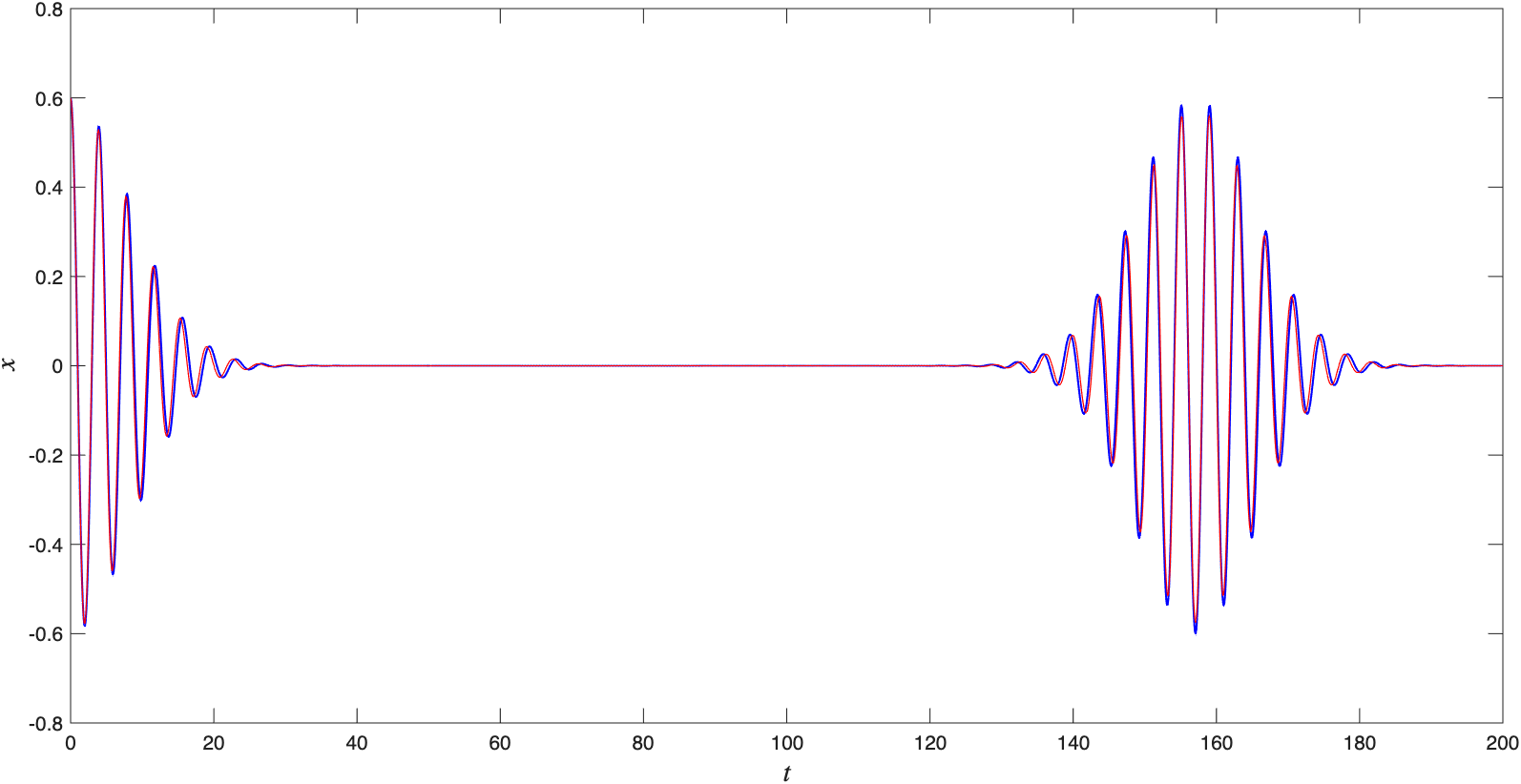}

  \caption{Comparison of the quantum (blue) and IMF (red) $x$-expectation values for spin 10 (top) and spin 50 (bottom) with $\kappa=0.5$, for an initial coherent state at  $z_0=0.8, \phi_0=0$.}
  \label{KerrExp}
\end{figure}

In the linear case, propagation via the IMF propagator yields exact results for arbitrary initial states. To gain insights into the performance of the approximation in the general case, it is instructive to consider just the Kerr-like interaction term in the Bose-Hubbard Hamiltonian, given by
\begin{equation}
    \hat{H}_{\rm Kerr} = \frac{2\kappa}{j}\hat{J}_z^2.
\end{equation}
The matrix elements of the time-evolution operator in the $\hat{J}_z$ basis are given as
\begin{align}
    U^{True}_{n,m}(t) &:= \langle j, n| \hat{U}^{True}(t) |j, m\rangle, \nn\\ &=\delta_{nm}e^{ -2\rmi\kappa t  m^2 / j }.
\end{align}
The quantum dynamics display the famous collapse and revival phenomena \cite{milburn1997quantum}, illustrated for an example in figures \ref{fig:kerrclassic} and \ref{KerrExp}. 

The mean-field equations of motion are directly integrated to yield
\begin{align}
    z(t) &=z_0, \\
    \phi(t) &= 4 \kappa z_0t + \phi_0, \\
    c(t) &= c_0 e^{\rmi 2j \kappa z(z-2)t}.
\end{align}
The mean-field trajectories move along circles of latitude, with frequency of rotation proportional to the value of $z_0$. Figure \ref{fig:kerrclassic} displays the mean-field trajectory in comparison to the expectation values of the full quantum dynamics for an initial coherent state with $z_0 =0.8, \phi_0 = 0$. The quantum expectation value shows the typical collapse behaviour due to the stretching of the quantum wave-packet, not accounted for in the individual mean-field trajectory, followed by a full revival at longer times (not shown in the figure) due to quantum interference. 

We will now demonstrate that while the IMF propagator is not exact for Kerr-interaction (even in the limit of $j\to\infty$), it does in fact recover the collapse and revival behaviour  well for even relatively small values of $j$ (such as $j=10$). 

The matrix elements of the IMF propagator in the $\hat{J}_z$ basis are given by 
\begin{align}
    U^{IMF}_{m,n}(t)   =& \frac{2j +1 }{4 \pi}\int^{2\pi}_{ 0}\int^1_{-1}  e^{\rmi 2j\kappa z(z-2)t}\langle j, m|z_0, 4\kappa z_0 t + \phi_0\rangle \langle z_0, \phi_0 | j, n \rangle \rmd z_0 \rmd \phi_0.
\end{align}
Using that
\begin{equation}
    \langle j , m | z_0, \phi_0 \rangle = \frac{1}{2^j}\sqrt{\binom{2j}{j-m}}\,(1+z_0)^{\frac{j+m}{2}} (1-z_0)^{\frac{j-m}{2}}\, e^{\rmi \phi_0(j-m)} , 
\end{equation}
and completing the square in the exponent yields 
\begin{align}
 \nn   U^{IMF}_{m,n}(t)
    &=\frac{2j +1 }{2 \pi}\sqrt{\binom{2j}{j-m}\binom{2j}{j-n}}e^{-\rmi 2\kappa t m^2 / j}  \times \int^{2\pi}_{ 0}e^{\rmi \phi_0(n-m)} \rmd \phi_0\\ & \times \int^1_{-1} 2^{-(2j+1)} (1+z_0)^{j+\frac{m+n}{2}} (1-z_0)^{j-\frac{m+n}{2}} e^{\rmi 2j t \kappa (z_0 - \frac{m}{j})^2} \rmd z_0. 
\end{align}
Performing the $\phi$ integration leads to the final expression 
\begin{align}
    U_{n,m}^{IMF} = \delta_{nm}e^{-\rmi 2\kappa t m^2 / j} g_m(2\kappa t),
\end{align}
where we have defined the function $g_m(A)$ is as
\begin{align}
    g_m(A) = \frac{(2j+1)!}{2^{2j+1}(j+m)!(j-m)!}\int^1_{-1} (1 + z_0)^{j+m}(1 - z_0)^{j-m}e^{\rmi  A j( z_0- \frac{m}{j})^2}  \rmd z_0.
    \label{equ-gma}
\end{align}
Importantly, the IMF propagator preserves the diagonal structure of the true propagator, however, there is an additional $m$-dependent factor.

\begin{figure}[tb]
  \centering
  \begin{minipage}[c]{0.65\textwidth}
    \centering
    \includegraphics[width=\textwidth]{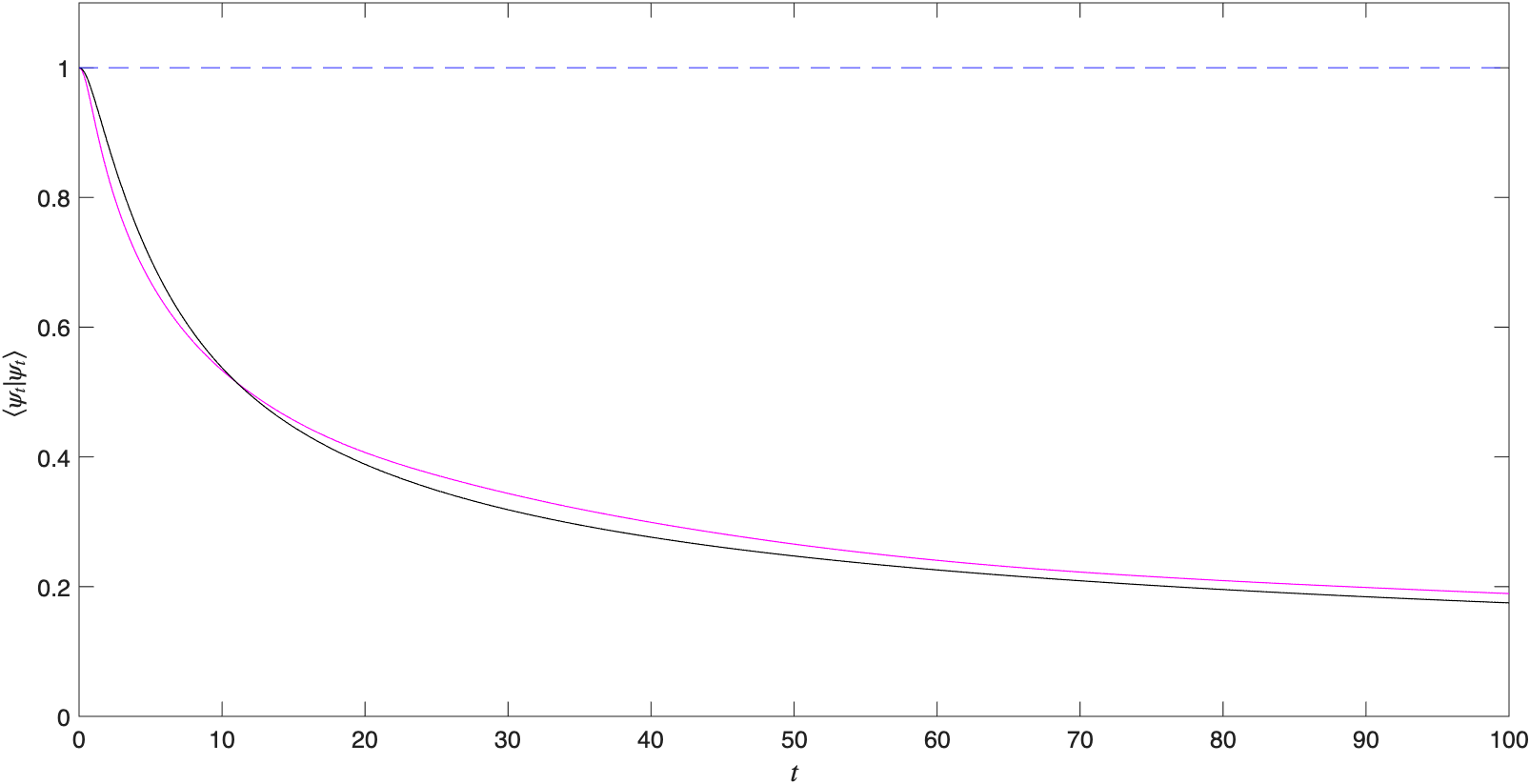}
  \end{minipage}

  \caption{The norms of the quantum state propagated with the IMF   for $\kappa=0.5$, for an initial coherent state at  $z_0=0.8, \phi_0=0$ are shown with spin 10 in magenta and spin 50 in black.}
  \label{KerrNormDecay}
\end{figure}

As a sanity check, we observe that in the case of vanishing interaction  ($\kappa=0$), the IMF propagator reduces to the identity matrix:
\begin{align}
    U^{IMF}_{m,n}|_{\kappa=0}     &=\delta_{nm} \frac{(2j+1)!}{2^{2j+1}(j+m)!(j-m)!}\int^1_{-1} (1 - z_0)^{j-m}(1 + z_0)^{j+m}  \rmd z_0, \nn \\
    &=\delta_{nm}\frac{2(2j+1)!}{(j+m)!(j-m)!}\frac{B(j+m+1, j-m+1)}{2}, \nn \\
    &= \delta_{nm} ,
\end{align}
where $B(x,y)$ denotes the beta-function and we have used that for integer $k,l$ it holds $B(k+1, l+1) = \frac{k!l!}{(k+l+1)!}$.

For $\kappa\neq 0$, the integral $g_m(2\kappa t)$ is not exactly solvable, but can be evaluated in a saddle-point approximation (see \ref{app-saddle} for details) in the semiclassical limit $j\to \infty$ to yield 
\begin{align}
    U^{IMF}_{n,m}(t; \text{large }j) &=  \frac{\delta_{nm} e^{-\rmi 2  \kappa t m^2 / j}}{\sqrt{1 -\rmi2 \kappa t (1-\frac{m^2}{j^2})}}. \label{UIMlargejapprox} 
\end{align}
Thus, we conclude that even in the $j\to \infty$ limit the IMF propagator is only an approximation to the true propagator with an $m$ dependent scaling factor. For small times or small interaction strength this factor is close to unity.

Note that the approximation is not unitary, resulting in non-physical norm decay. This is a feature that is commonly found in semi-classical propagators, and in flat space can often be mitigated by Herman-Kluk type generalisations \cite{ray2016dynamics,heller2018semiclassical} or by an `on the fly' renormalisation technique \cite{tatchen2011renormalization}. 

\begin{figure}[tb]
  \centering
  \includegraphics[width=0.23\textwidth]{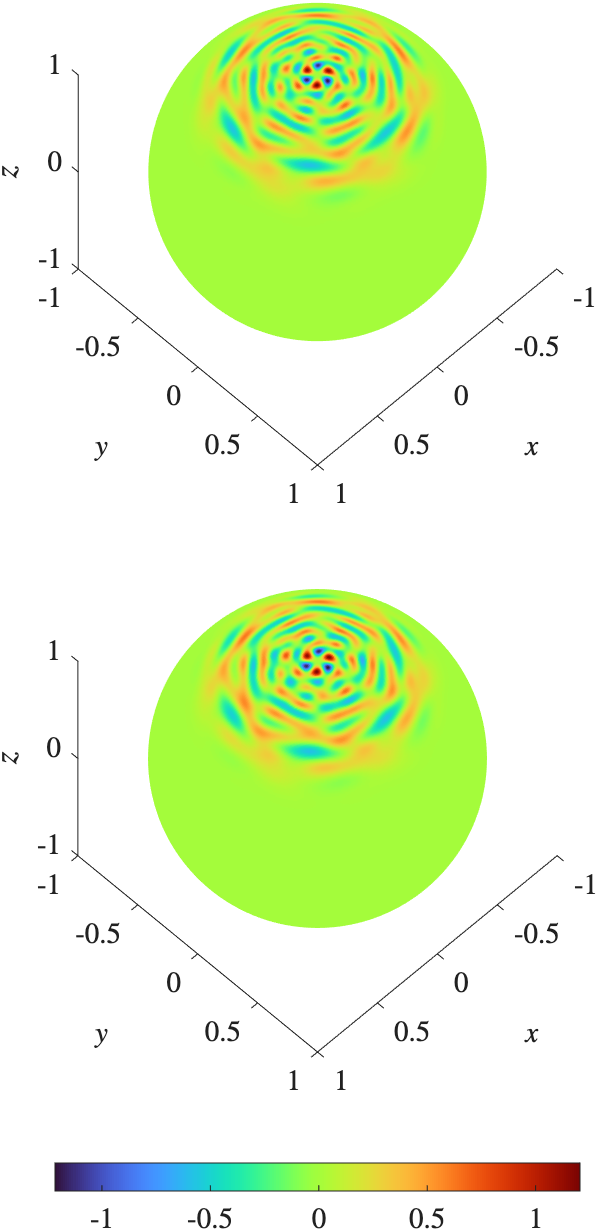}
  \hspace{0.3em}
  \includegraphics[width=0.23\textwidth]{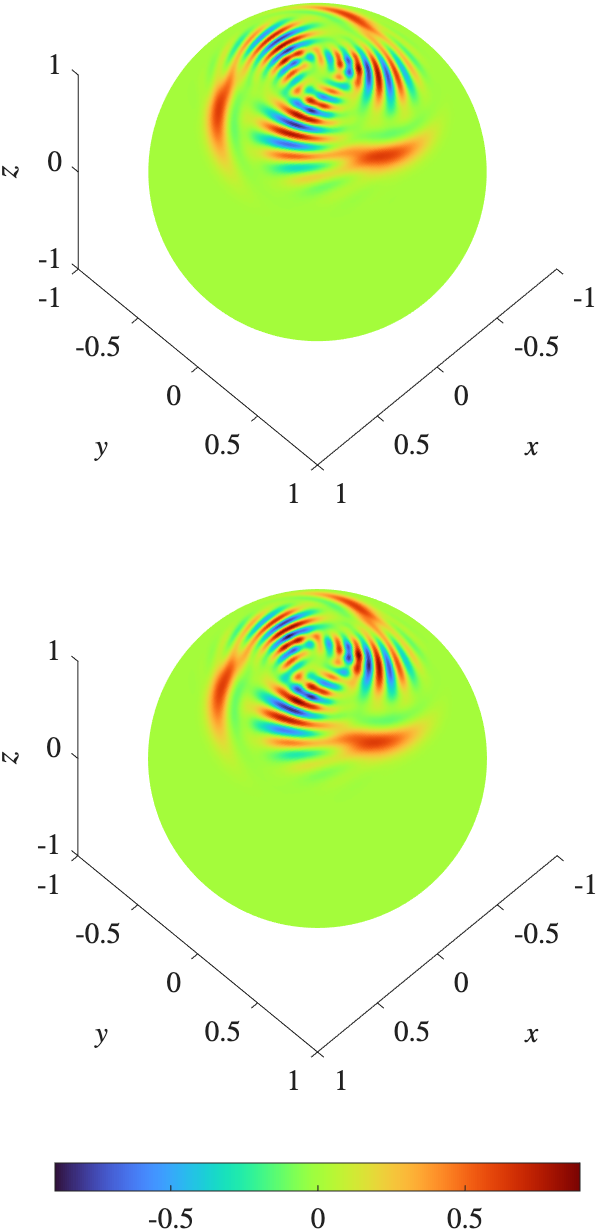}
  \hspace{0.3em}
  \includegraphics[width=0.23\textwidth]{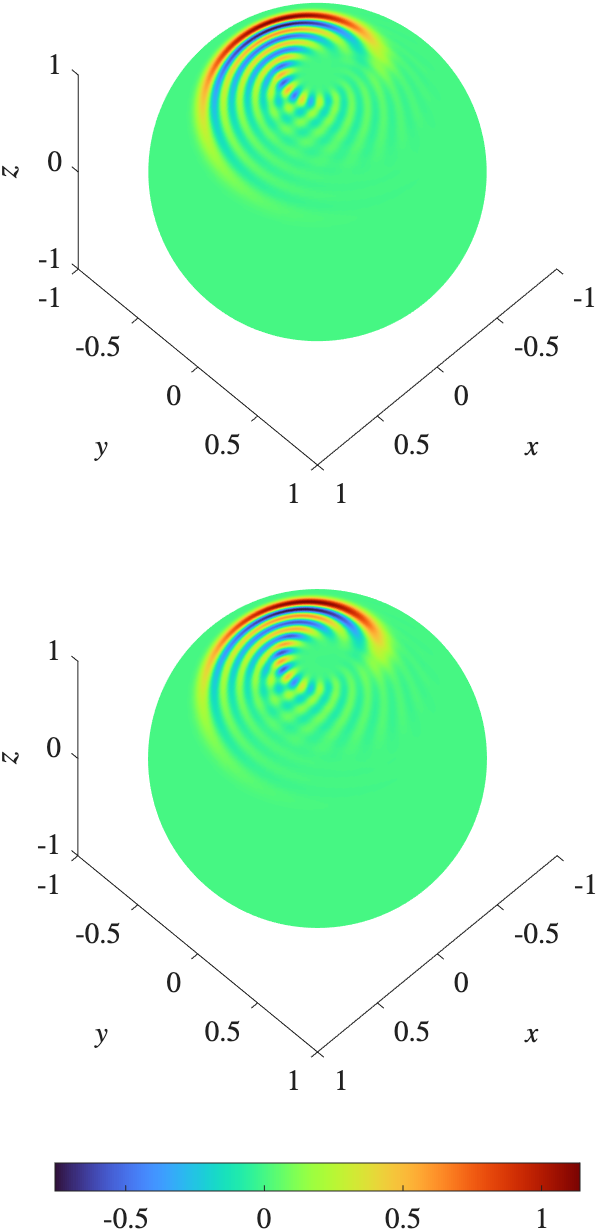}
  \hspace{0.3em}
  \includegraphics[width=0.23\textwidth]{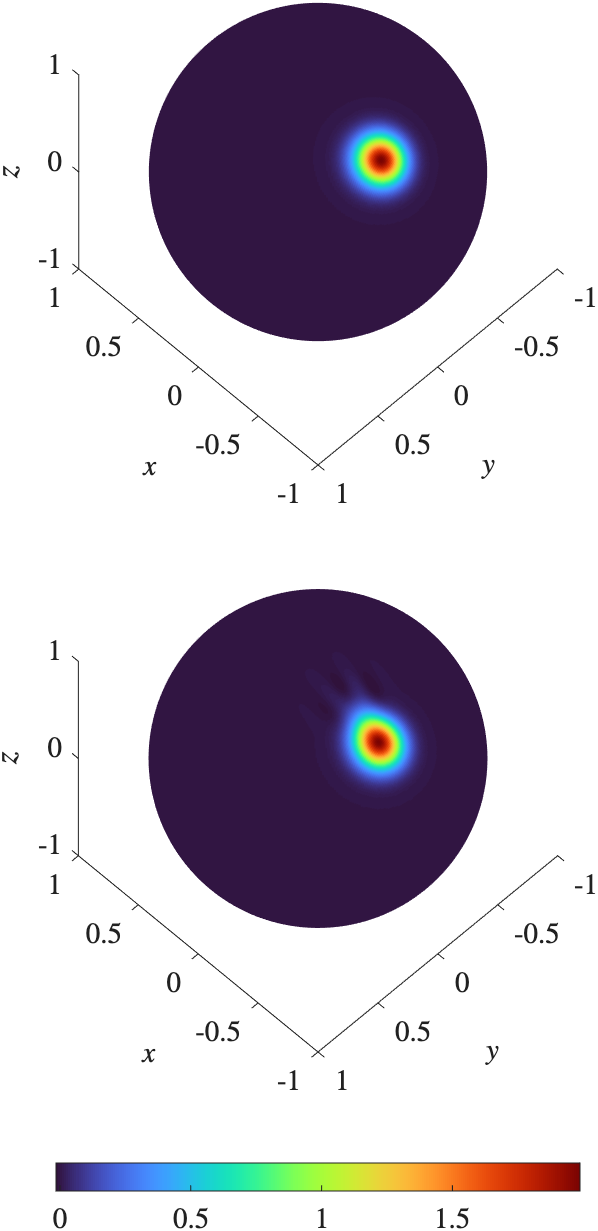}
  
  \caption{Comparison of the quantum (top) and IMF (bottom) Wigner functions for a spin 50 system with parameters $\kappa=0.5, z_0=0.8, \phi_0=0$. From left to right, the system is depicted at $t=25$, $t=50$, $t=150$ and $t=157$.}
  \label{KerrWigners}
\end{figure}

Despite this discrepancy to the full propagator, the IMF propagator often leads to surprisingly accurate results. This is illustrated in figure \ref{KerrExp} which shows the $x$-expectation values obtained from the IMF propagator compared to the exact quantum dynamics, for $\kappa=0.5$ and an initial coherent state centred at $z_0=0.8, \phi_0=0$ for two values of $j$. Figure \ref{KerrNormDecay} shows the norm of the IMF state, which decays rapidly. Despite this unphysical norm decay, we observe a good qualitative agreement for the expectation values for both values of $j$, in particular for the higher spin. Figure \ref{KerrWigners} shows the normalised Wigner functions (see \ref{sec-appendix-wigner})  for the spin 50 system at four times. The Wigner functions generated by IMF propagator recover even fine structures of the quantum dynamics and are nearly indistinguishable from their exact counterparts.

\subsection{Numerical Implementation of the IMF propagator for the Bose-Hubbard dimer}

In the general Bose-Hubbard case, for non-zero $\nu$ and $\kappa$, the IMF needs to be implemented numerically. This includes two numerical approximations, one in the solution to the equations of motion, and the second in approximating the integral (\ref{U_IM_general}). In the following we use the inbuilt ode45 solver in MATLAB, to numerically solve (\ref{BHEOMphi_limit})-(\ref{BHEOMc_limit}). We discretise the problem  using a  Gauss-Legendre quadrature, which is easily implemented and efficient for approximating integrals on the sphere. In the $\phi$ direction we use equidistant spacing, and in the $z$ direction the grid points are the roots of Legendre polynomials, with appropriate weighting factors (see \cite{abramowitz1948handbook}, section 25.4.29).  The size of the grid will affect the accuracy of the approximation and a larger grid is needed for larger values of $j$, and in general for long time propagation a denser grid is needed to converge.

\begin{figure}[tb]
    \centering
    \vspace{0.5em}
    \includegraphics[width=0.45\linewidth]{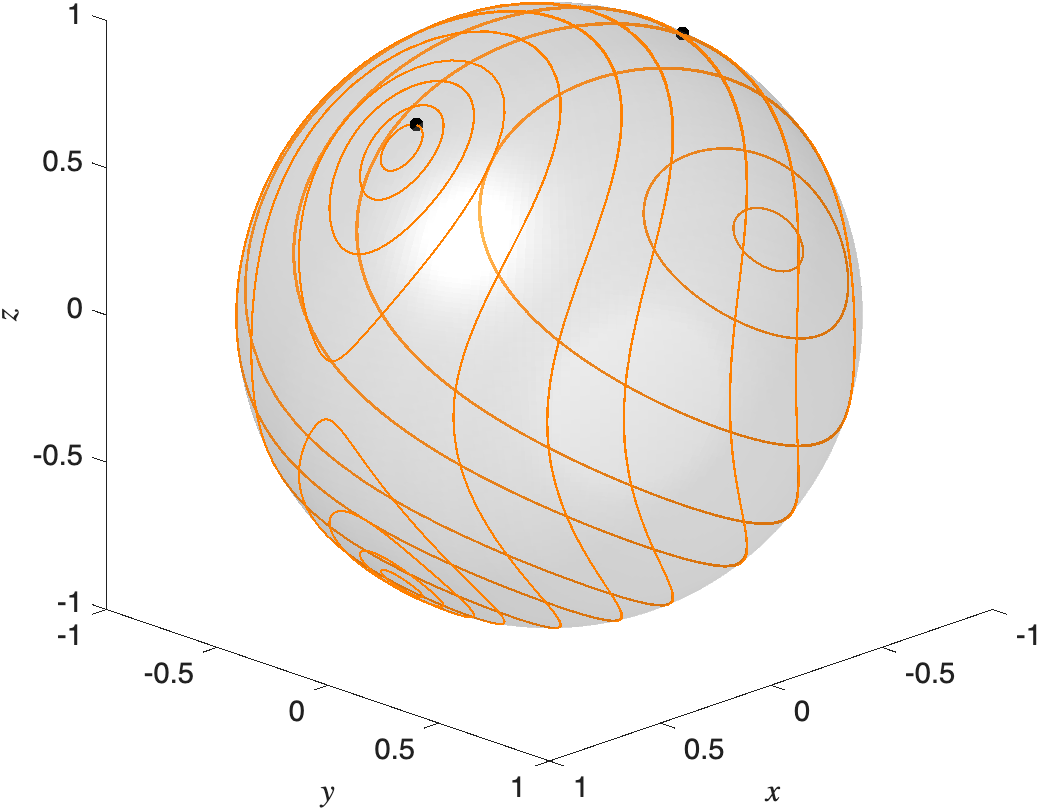}
    \caption{Mean field trajectories (\ref{BHEOMphi_limit}) and (\ref{BHEOMz_limit}) for the Bose-Hubbard system, with $\epsilon=0$, $\nu = 1$, and  $\kappa =0.75$. The black dots indicate the points with $z=0.8, \phi = 0,\pi$. }
    \label{BH_meanfield_traj}
\end{figure}

As an example we consider the Bose-Hubbard dynamics for the parameter values $j=10, \varepsilon=0, \nu=1, \kappa=0.75$. The corresponding mean-field phase space portrait is depicted in figure \ref{BH_meanfield_traj}. The mean-field dynamic has four fixed points; one elliptic fixed point  located at $z=0$ and $\phi=\pi$ (at the back of the sphere), a hyperbolic fixed point at $z=0$, and $\phi=0$, and a pair of elliptic self-trapping fixed points located at 
\begin{equation}
    z_{\rm st}=\pm\sqrt{1-\left(\frac{\nu}{2\kappa}\right)^2} \quad {\rm and} \quad \phi_{\rm st}=0.
\end{equation}
An initial coherent state in the neighbourhood of the single elliptic fixed point at $\phi=\pi$ will perform a variation of the collapse and revival phenomenon observed in the pure interaction case. However, the presence of the hyperbolic fixed point resulting from the competition of the interaction $\hat J_z^2$ and the hopping term $\hat J_x$, lead to incomplete revivals. 

\begin{figure}[tb]
    \centering
    \includegraphics[width=0.65\textwidth]{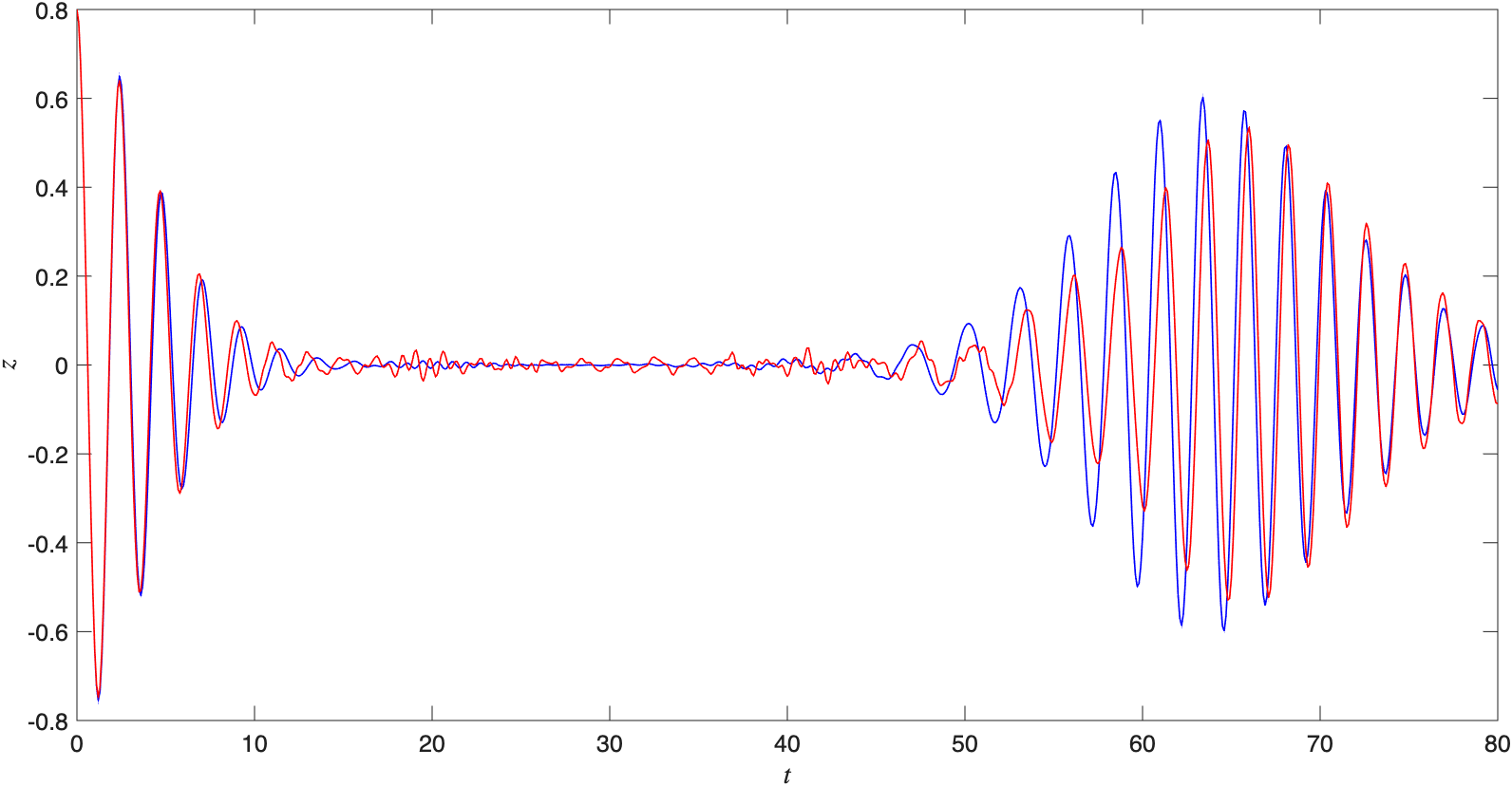}
    
    \vspace{1em}
    
    \includegraphics[width=0.2\textwidth]{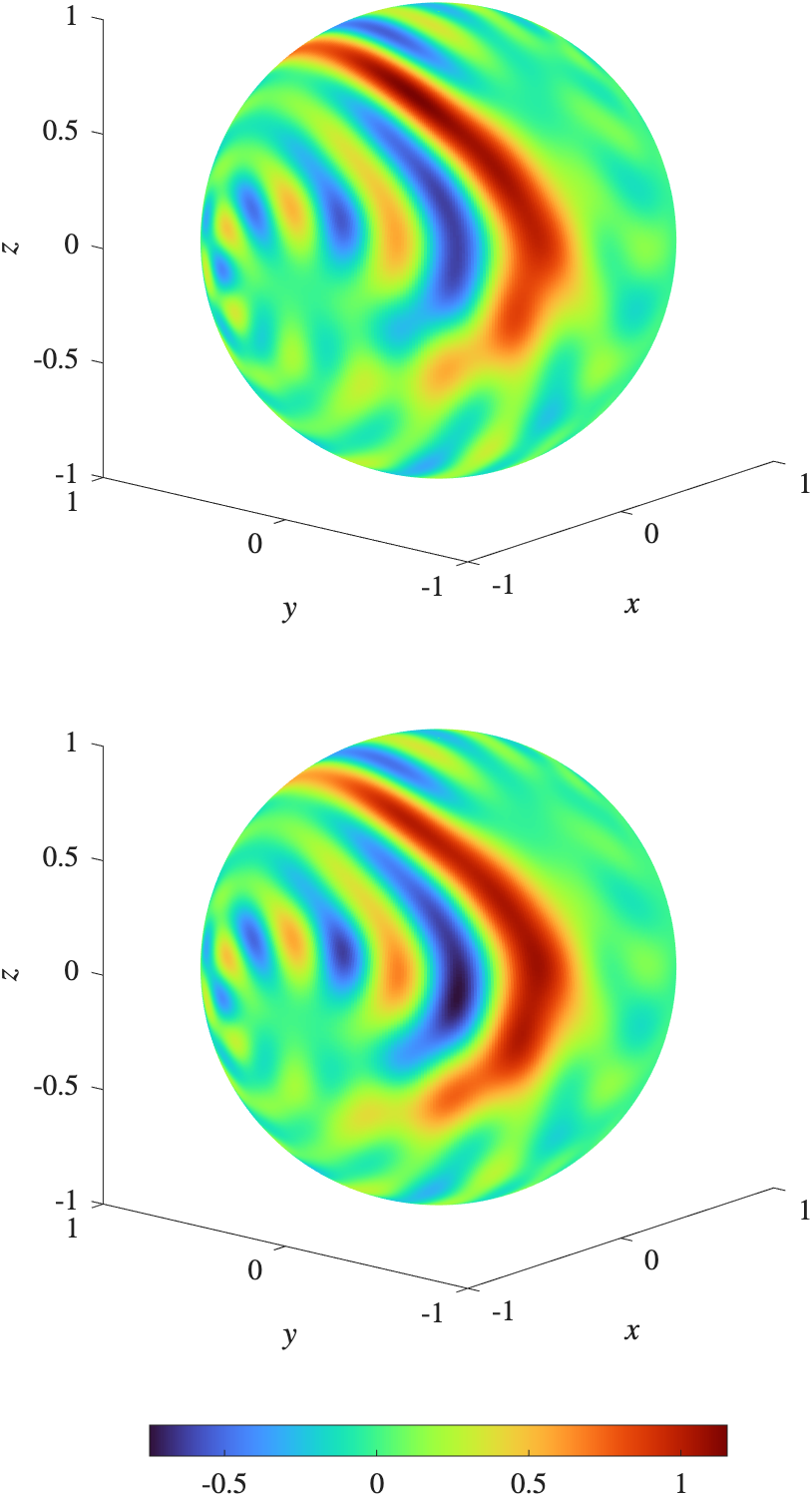}
    \includegraphics[width=0.2\textwidth]{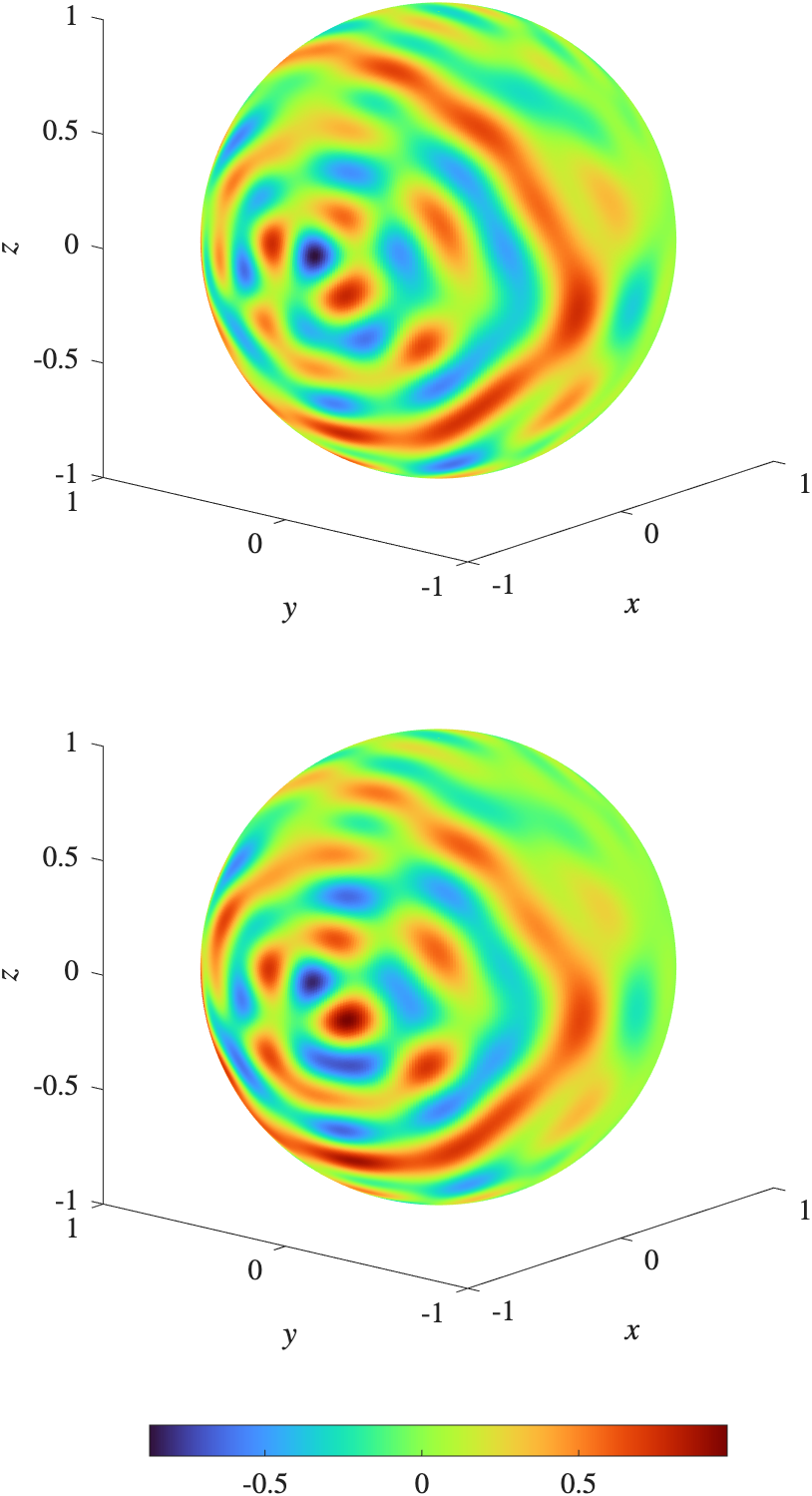}
    \includegraphics[width=0.2\textwidth]{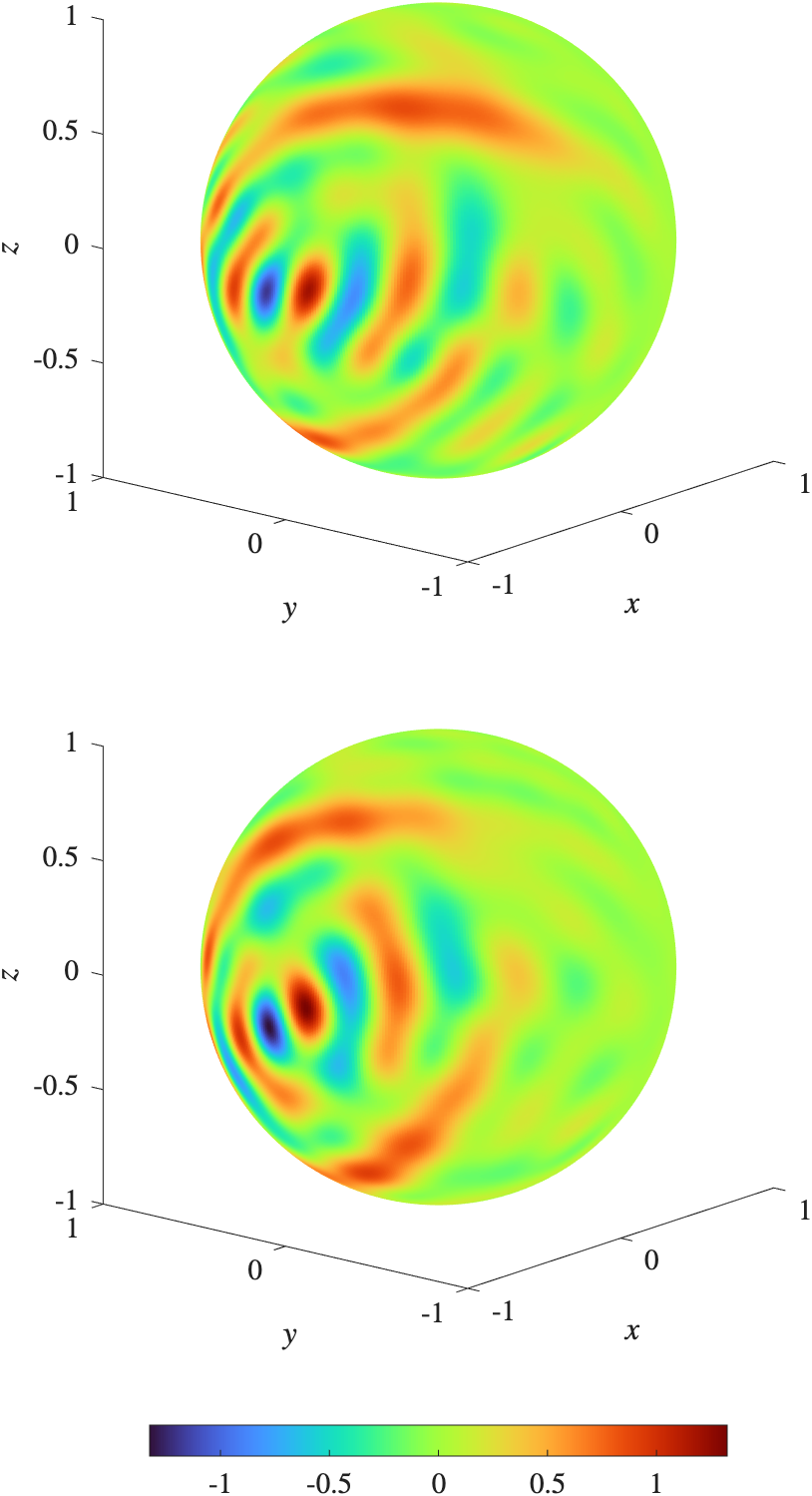}
    \includegraphics[width=0.2\textwidth]{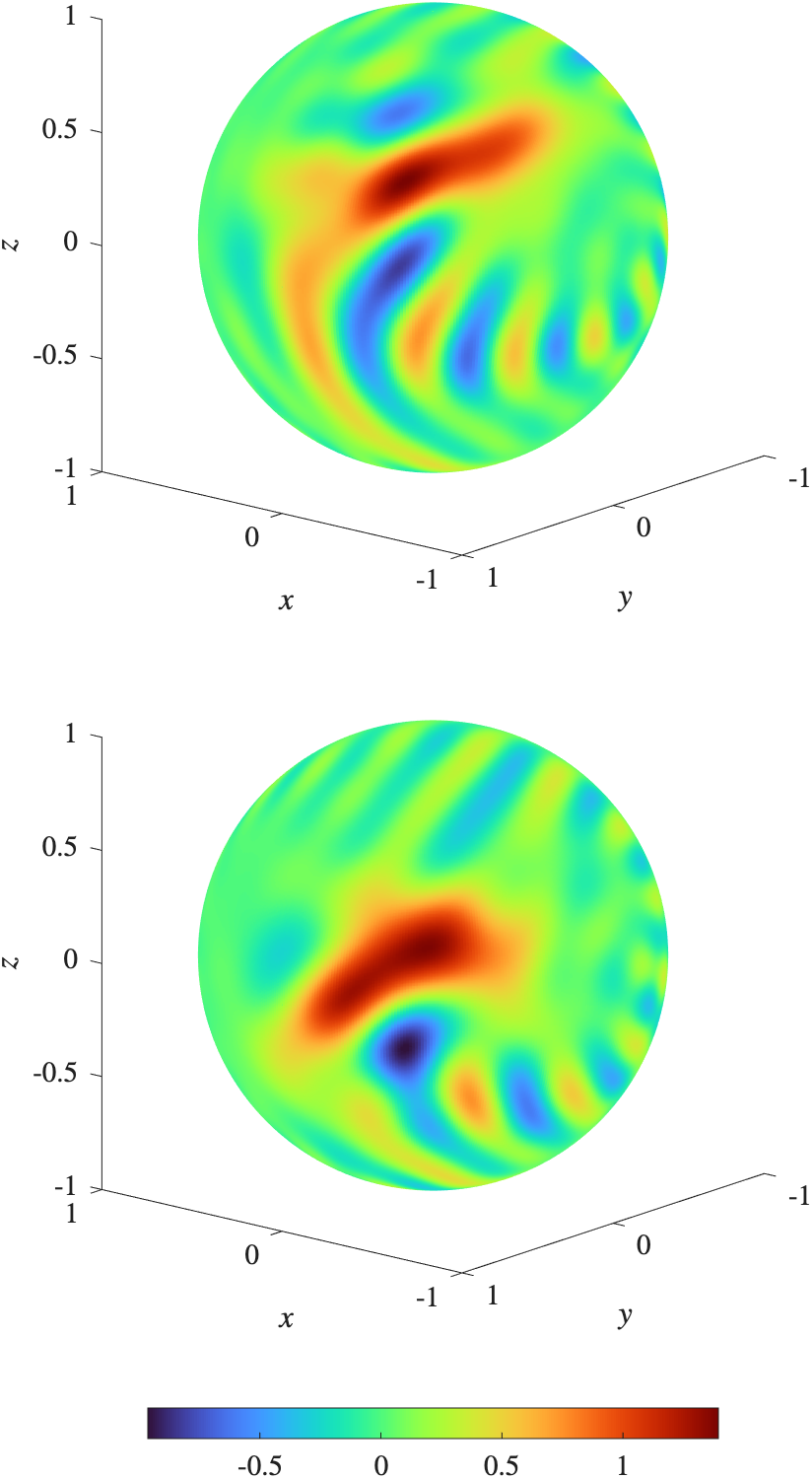}
  \caption{Numerically implemented IMF and exact many-particle evolution for a spin 10 Bose-Hubbard dimer with parameters $\epsilon=0,\nu=1, \kappa=0.75,$ for an initial coherent state at $ z_0 = 0.8, \phi_0=\pi$. The $z-$expectation value is shown in the top figure (blue line quantum, red line IMF approximation) as well as the Wigner functions for the true many-particle (middle) and IMF (bottom) propagations at times $t=5,\, 10,\, 30$  and $70$. }
  \label{BHB&RexpandWigner}
\end{figure}

In figure \ref{BHB&RexpandWigner} we show the $z$-expectation values over time, and the Wigner functions for selected times for both the full quantum dynamics and the numerically implemented IMF state, for an initially coherent state with $z_0=0.8, \phi=\pi$, located on a mean-field orbit around the elliptic fixed point. A grid size of $500 \times 500$ was used for the approximation. We observe a good qualitative agreement in the $z$-expectation value as well as the Wigner functions at a variety of times between the numerical IMF propagation and the full many-body dynamics, though the IMF results are not as close to the full dynamics as in the pure interaction case. This is most likely due to the presence of the hyperbolic fixed point in the mean-field dynamics, and a stronger sensitivity to phase coherence of the full many-particle dynamics.  While we have verified that there is only a negligible difference to an IMF simulation carried out with a grid size of $300\times 300$, it is possible that the approximation could be improved by using a much larger grid size and/or tighter tolerances in the solution of the individual ODEs.

\begin{figure}[tb]
  \centering
  \includegraphics[width=0.65\textwidth]{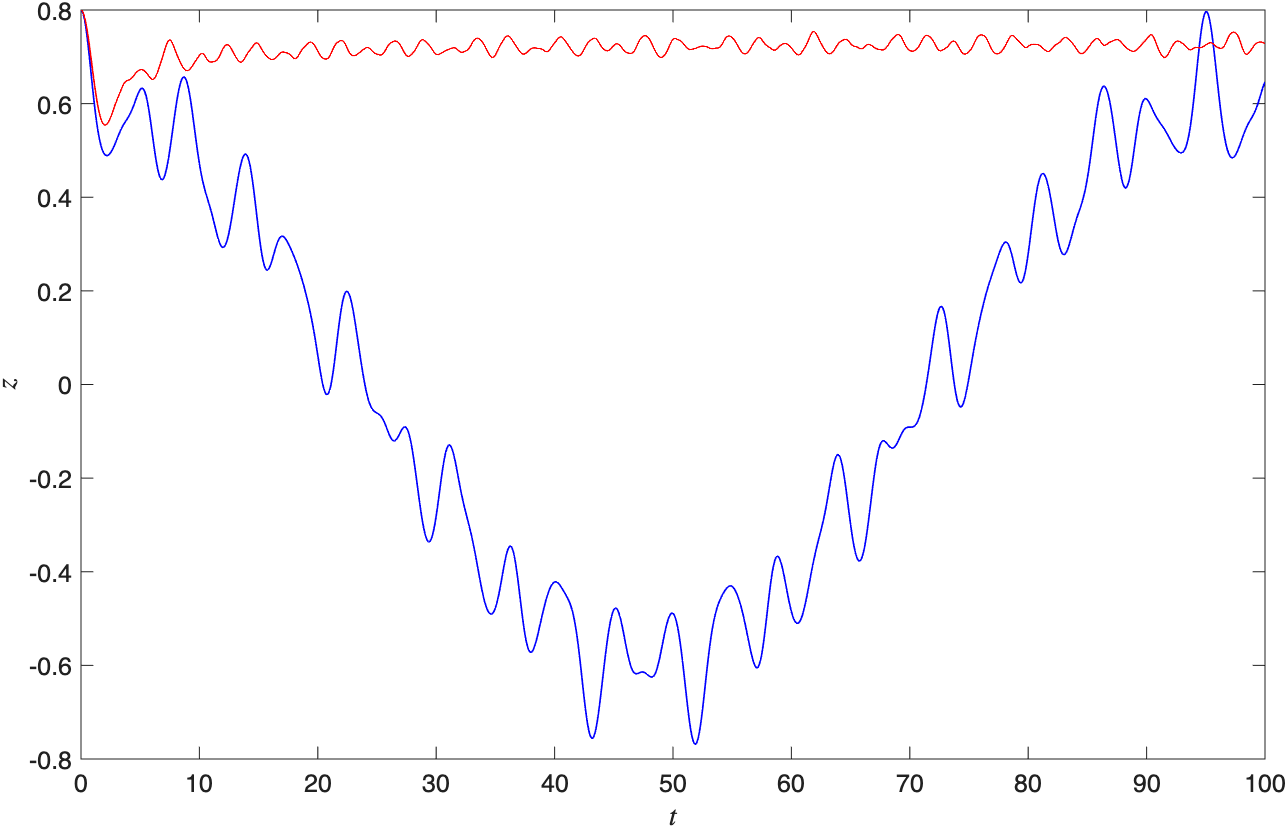}
      
    \vspace{1em}
    
    \includegraphics[width=0.23\textwidth]{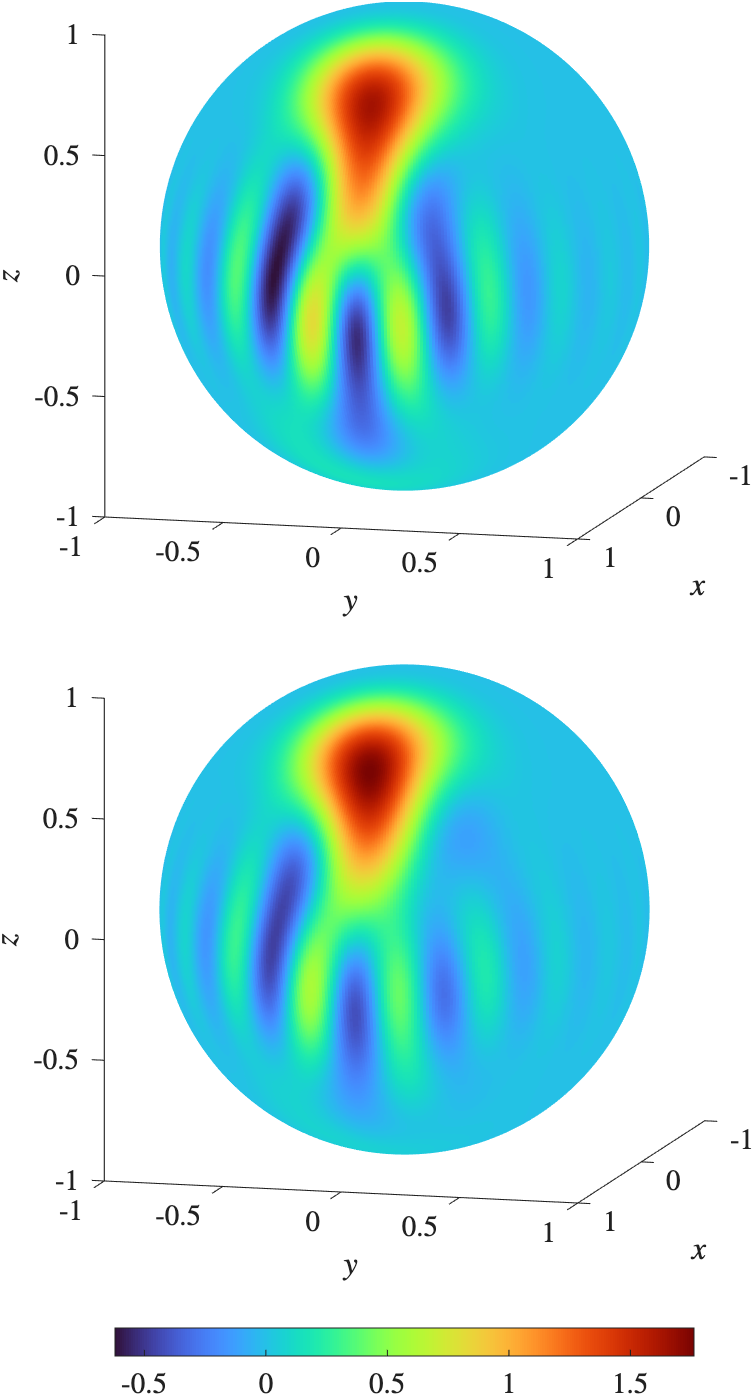}
    \includegraphics[width=0.23\textwidth]{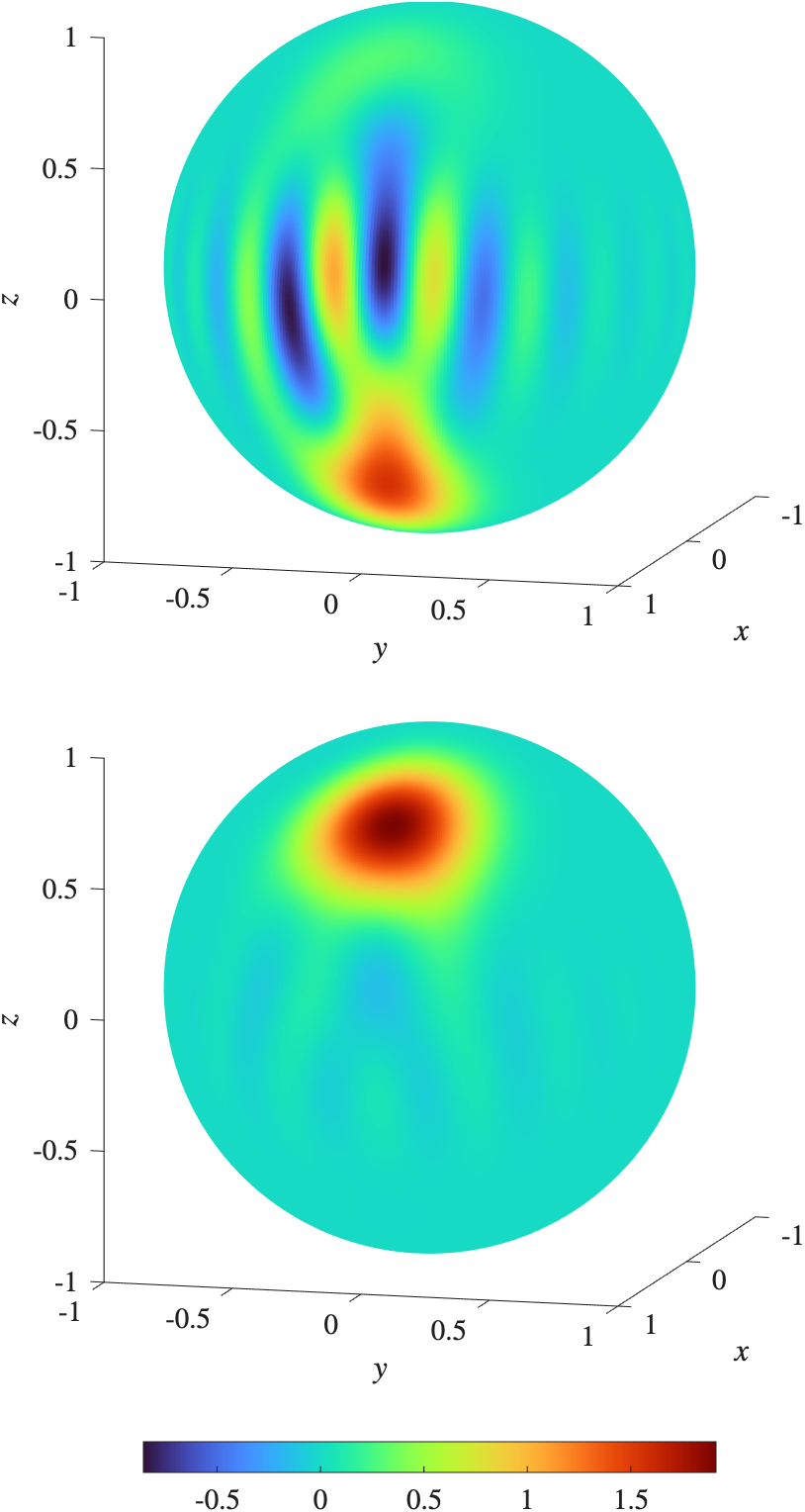}
    \includegraphics[width=0.23\textwidth]{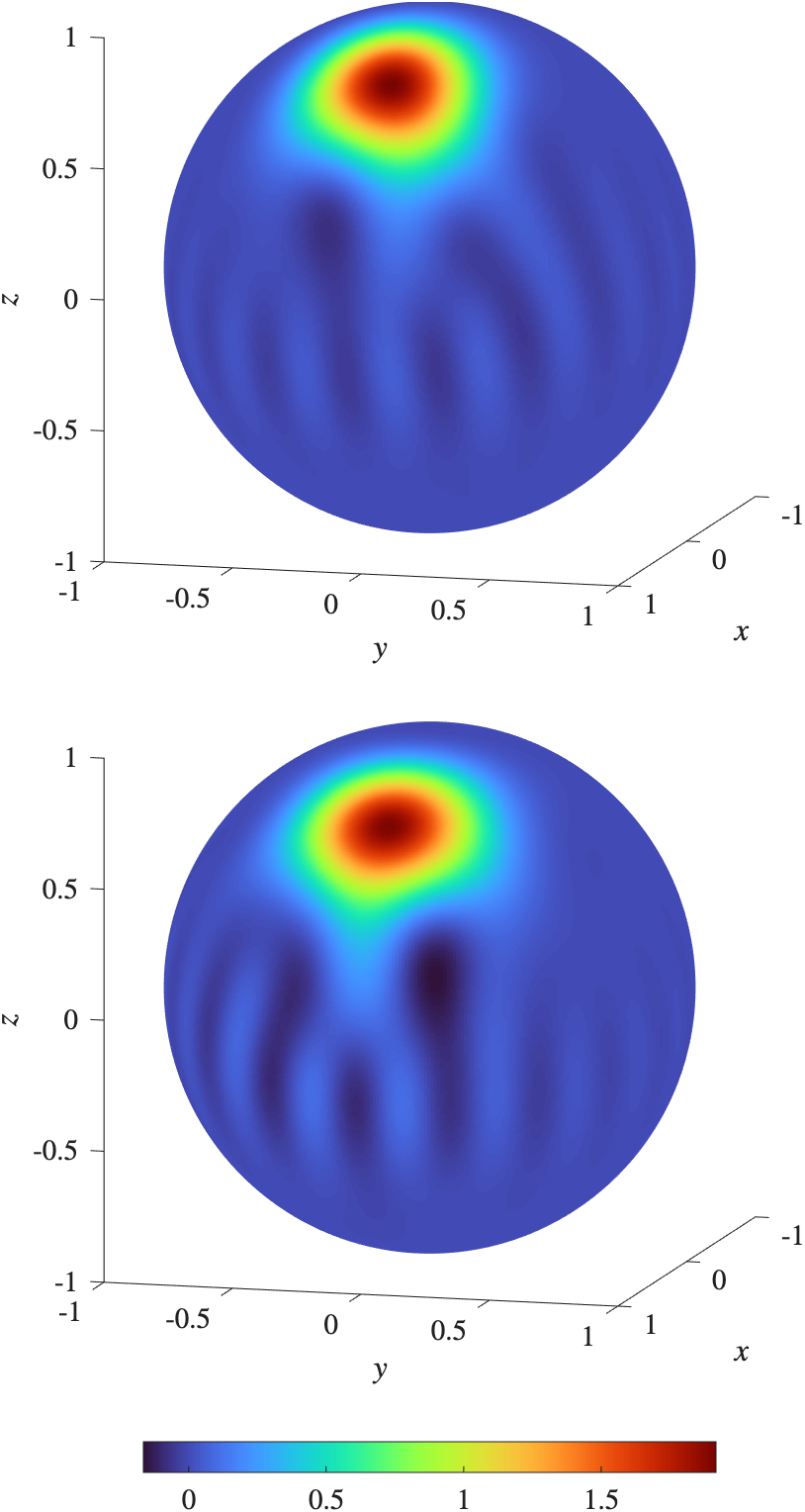}
  \caption{Numerically implemented IMF and exact many-particle evolution for a spin 10 Bose-Hubbard dimer with parameters $\epsilon=0,\nu=1, \kappa=0.75,$ for an initial coherent state at $ z_0 = 0.8, \phi_0=0$. The $z$-expectation value is shown in the left figure (blue line quantum, red line IMF approximation) as well as the Wigner functions for the true many-particle (top row on the right) and IMF (bottom row on the right) propagations at times $t=3,\, 40,$  and $95$.}
  \label{BHTunnellingexpandWigner}
\end{figure}

In figure \ref{BHTunnellingexpandWigner} we show results of the time-evolution with the IMF propagator compared to the exact many-particle dynamics for an initial coherent state centred at $z_0=0.8, \phi_0=0$. This is close to one of the mean-field self-trapping states and experiences many-particle tunnelling to the other self-trapping state. Like many other initial value semiclassical propagators, the IMF propagator is not able to reproduce the tunnelling, although looking at the Wigner functions after a tunnel period, it in fact recovers some of the features of the full dynamics even for long times. In the next section we will demonstrate that a time-sliced version of the IMF leads to vast improvements that accurately reproduce even the tunnelling behaviour.

\section{A Time-Slicing Initial Value Coherent State Propagator}
\label{sec-TS}
Though the IMF propagator is not exact in the non-linear case, it may be argued that for short times, on which the individual coherent states do not spread significantly, it is a good approximation. This motivates the use of a time-slicing method, similar to that of Huber's and Heller's Hybrid Mechanics in flat space \cite{huber1988hybrid,davis1984comparisons}, where the propagator is applied for a short time before re-expanding the resulting state in the original coherent-state basis and then repeating the procedure. This approach has been shown to be able to describe barrier tunnelling \cite{ grossmann2000semiclassical,bergold2022error}. 

The IMF propagator transports each component along its mean-field trajectory and thus cannot by itself generate population transfer between dynamically disconnected classical mean-field regions, such as the two self-trapping islands. A re-expansion into the coherent-state basis after a short propagation, however, includes small amplitudes on nearby coherent states lying closer to or on the other side of the separatrix. Repeating the short-time propagation and re-expansion allows these small amplitudes to accumulate coherently over many steps, yielding tunnelling behaviour on long time scales. We shall show in what follows that adapting our IMF propagator in this way allows us to recover even the many-body tunnelling dynamics between mean-field self-trapping states with astonishing accuracy for a wide range of particle numbers and time scales. 

Beginning at time zero, we evolve the initial state up to time $\delta t$ using the IMF propagator, and take the result as our new initial state, resetting the time to zero. We repeat this process a further $M-1$ times, until the desired final time $t=M\delta t$, to find our final approximate time evolved state. Thus, the time-sliced IMF time-evolution  operator can be expressed as
\begin{equation}
    \hat{U}^{TS}(t) =  \prod^M_{m=1} \hat{U}^{IMF}( \delta t). \label{TSpropexpression}
\end{equation}
If the grid is large enough, in principle the smaller one chooses $\delta t$ the better the result, though there is a trade off with an accumulated error from the repeated re-expansion on a finite grid. In practice, however, we have found that choosing a smaller $\delta t$ allows for a comparatively smaller grid size, as less coherent states are needed to approximate the dynamics for short times. Further,  we only need to solve the individual ODEs once for the first short time interval, which is much shorter than the time necessary for the single-slice IMF implementation. This leads to a significant speed up when moving to a time-sliced version with small $\delta t$.  For our numerical calculations we picked $\delta t$ such that the norm does not drop below a threshold of 0.9999. Note that the rate of the norm decay in each segment depends on the initial condition. We observe a convergence of the resulting dynamics with decreasing value of $\delta t$.  However, it should be noted that the limit of $\delta t\to 0$ does not necessarily yield the exact many-particle dynamics. 

\begin{figure}[tb]
  \centering
  \includegraphics[width=0.49\textwidth]{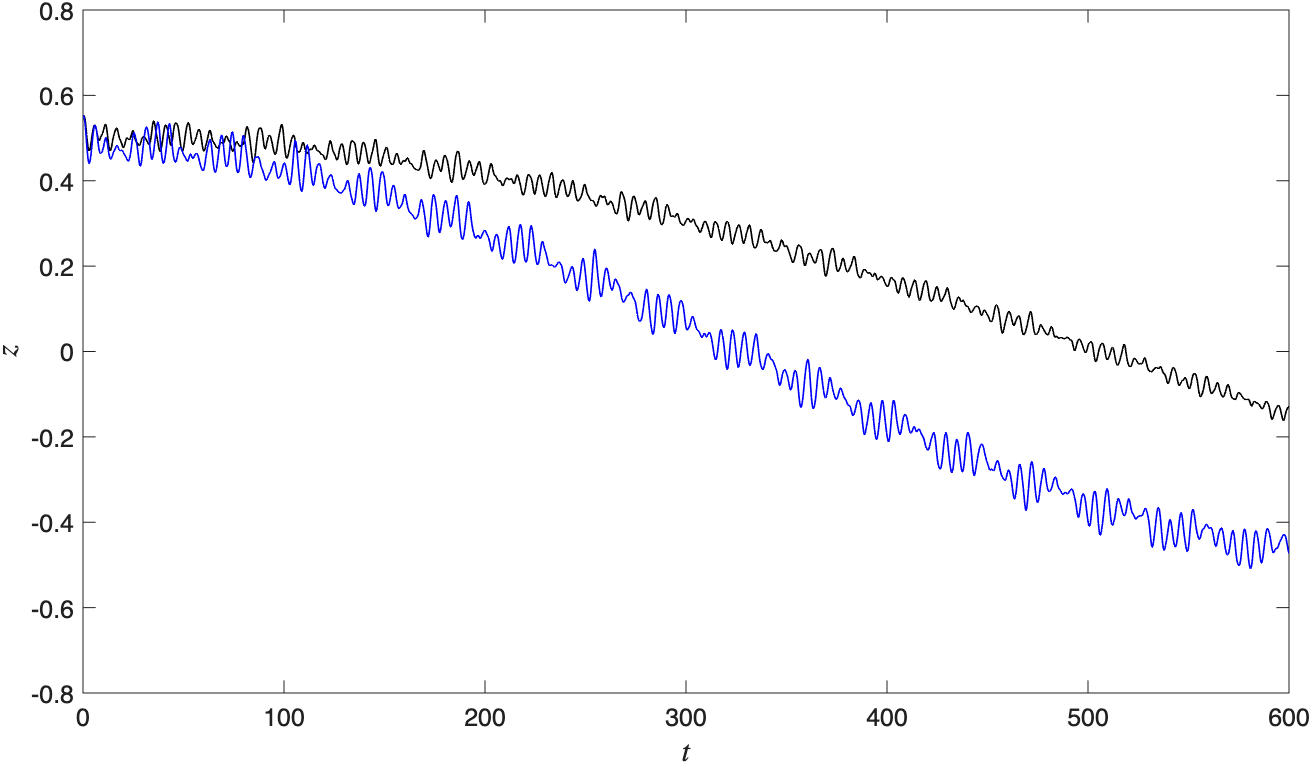}
  \includegraphics[width=0.49\textwidth]{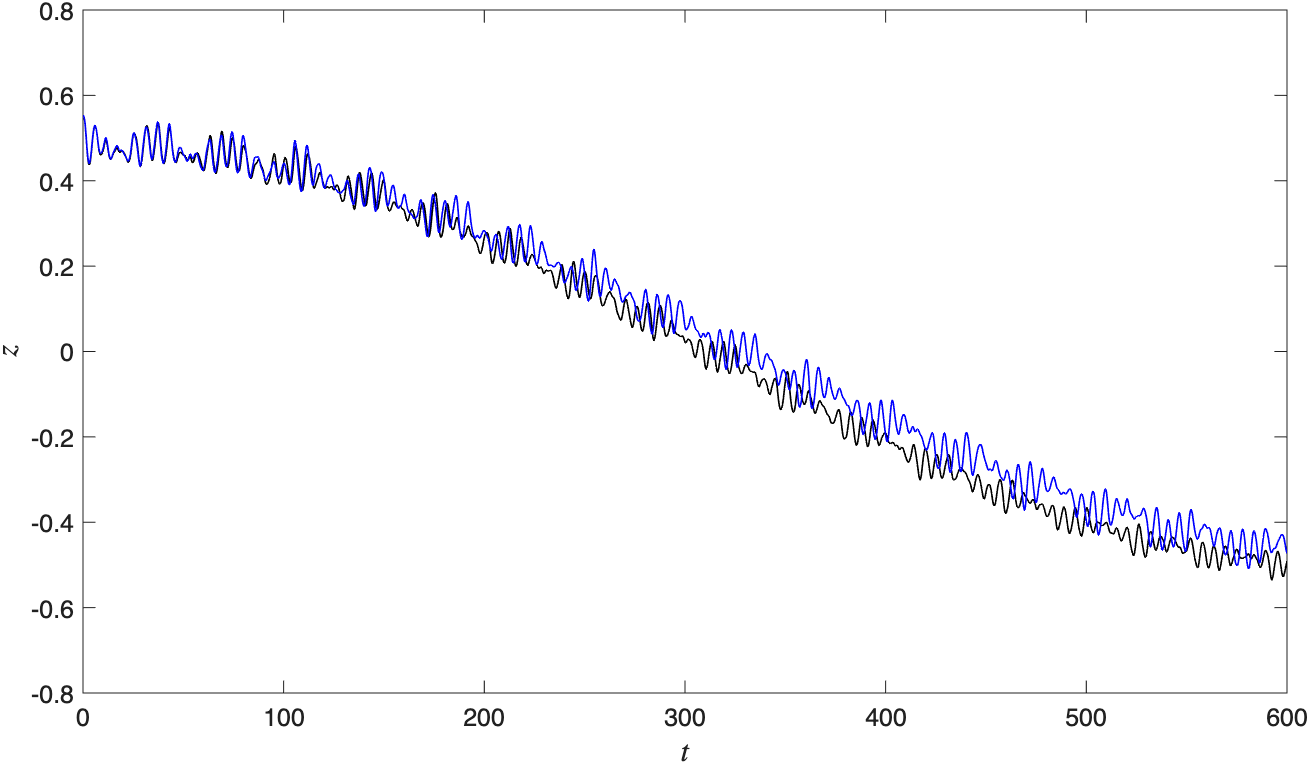}
  
  \caption{Comparison of quantum (blue) and TS-IMF (black) $z$-expectation values for a spin 50 Bose-Hubbard system, $\varepsilon=0, \nu=1, \kappa=0.6$, and an initial coherent state at  $z_0=0.553$ and $\phi_0=0$, for two different scaling parameters in the TS-IMF propagator ($1$ on the left, and $1-\frac{1}{2j+1}=0.99$ on the right). The TS-IMF was simulated with a grid size of $150\times 150$ and $\delta t=0.001$. }
  \label{BHtimesliceExp_unscaled vs predicted}
\end{figure}

\begin{figure}[tb]
  \centering
  \includegraphics[width=0.65\textwidth]{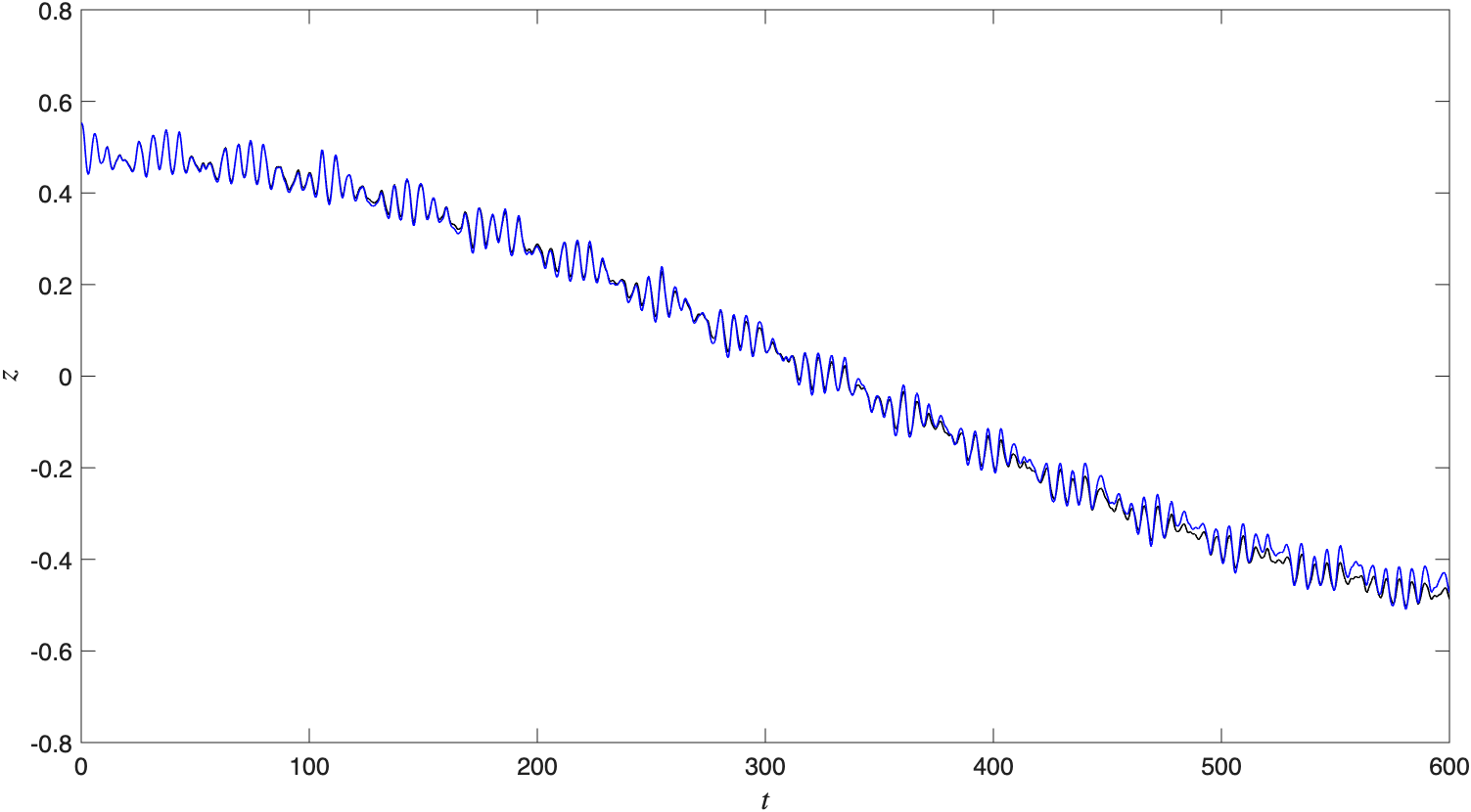}
  \vspace{0.5em}\\
  \includegraphics[width=0.23\textwidth]{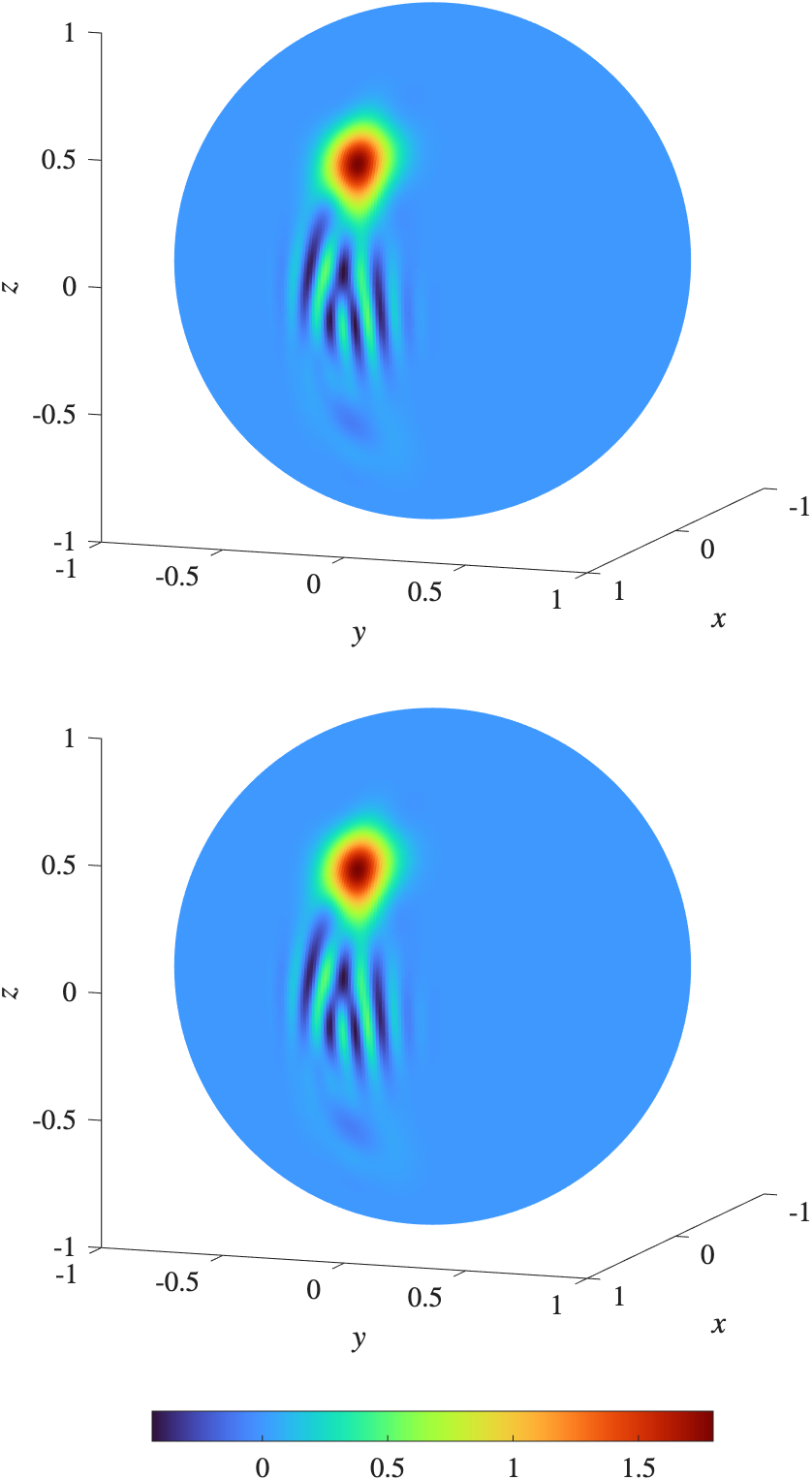}
  \hspace{0.1em}
  \includegraphics[width=0.23\textwidth]{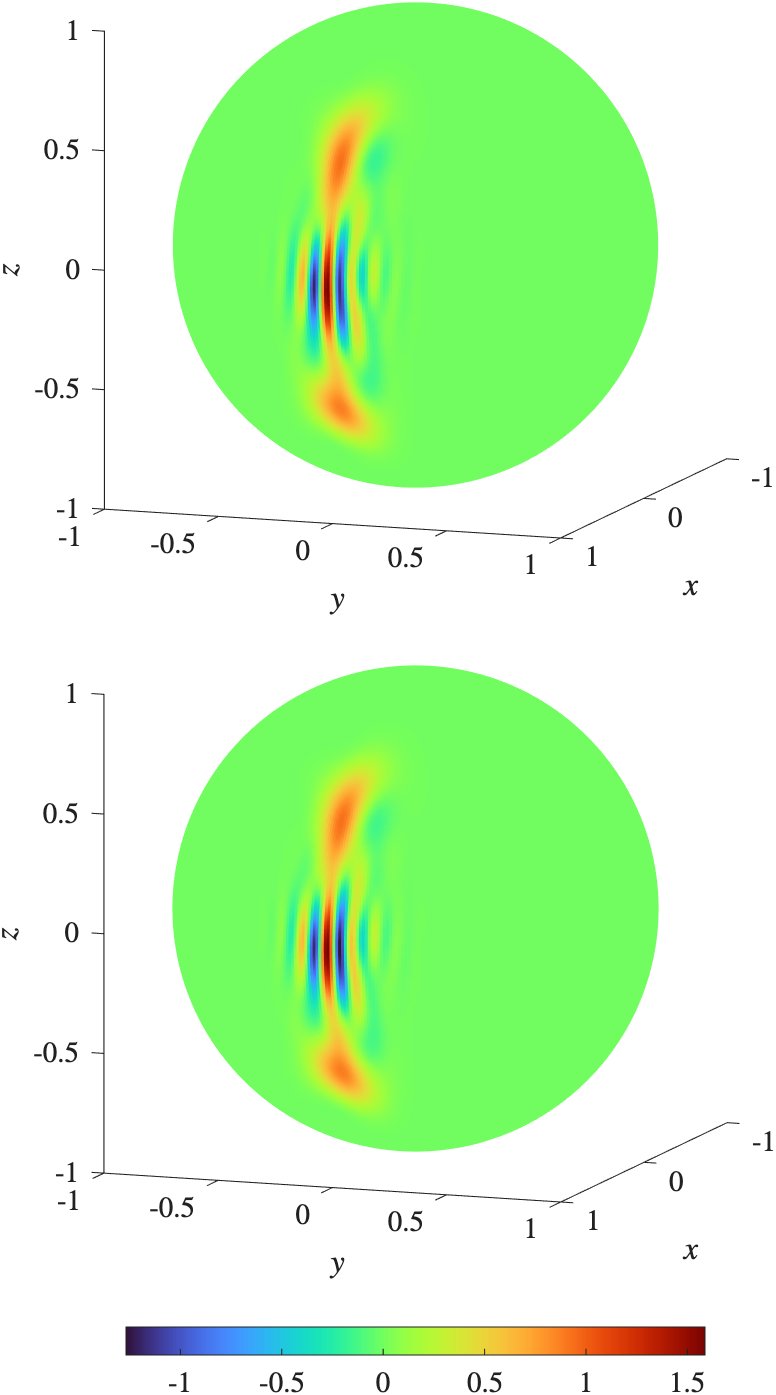}
  \hspace{0.1em}
  \includegraphics[width=0.23\textwidth]{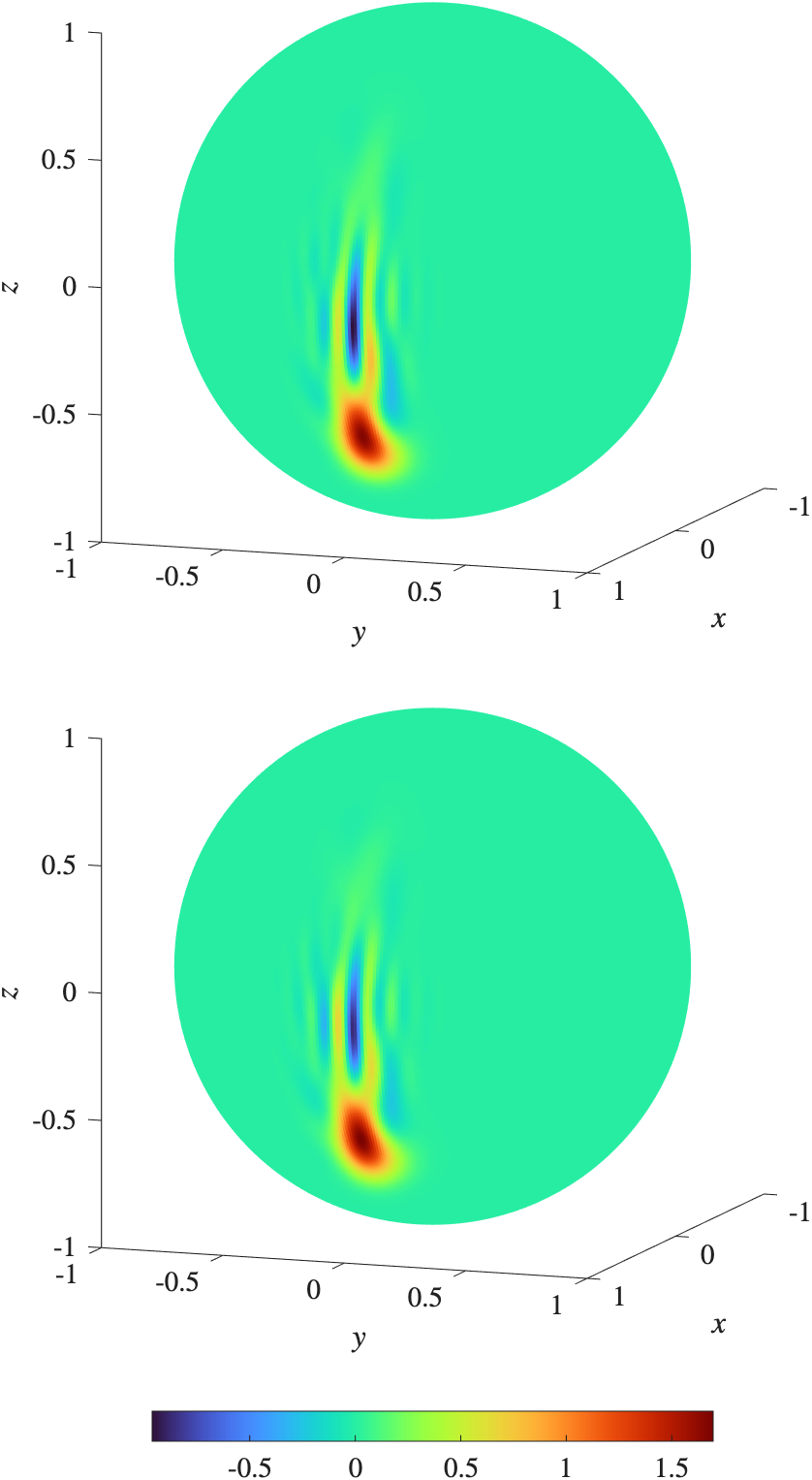}
  \hspace{0.1em}
  \includegraphics[width=0.23\textwidth]{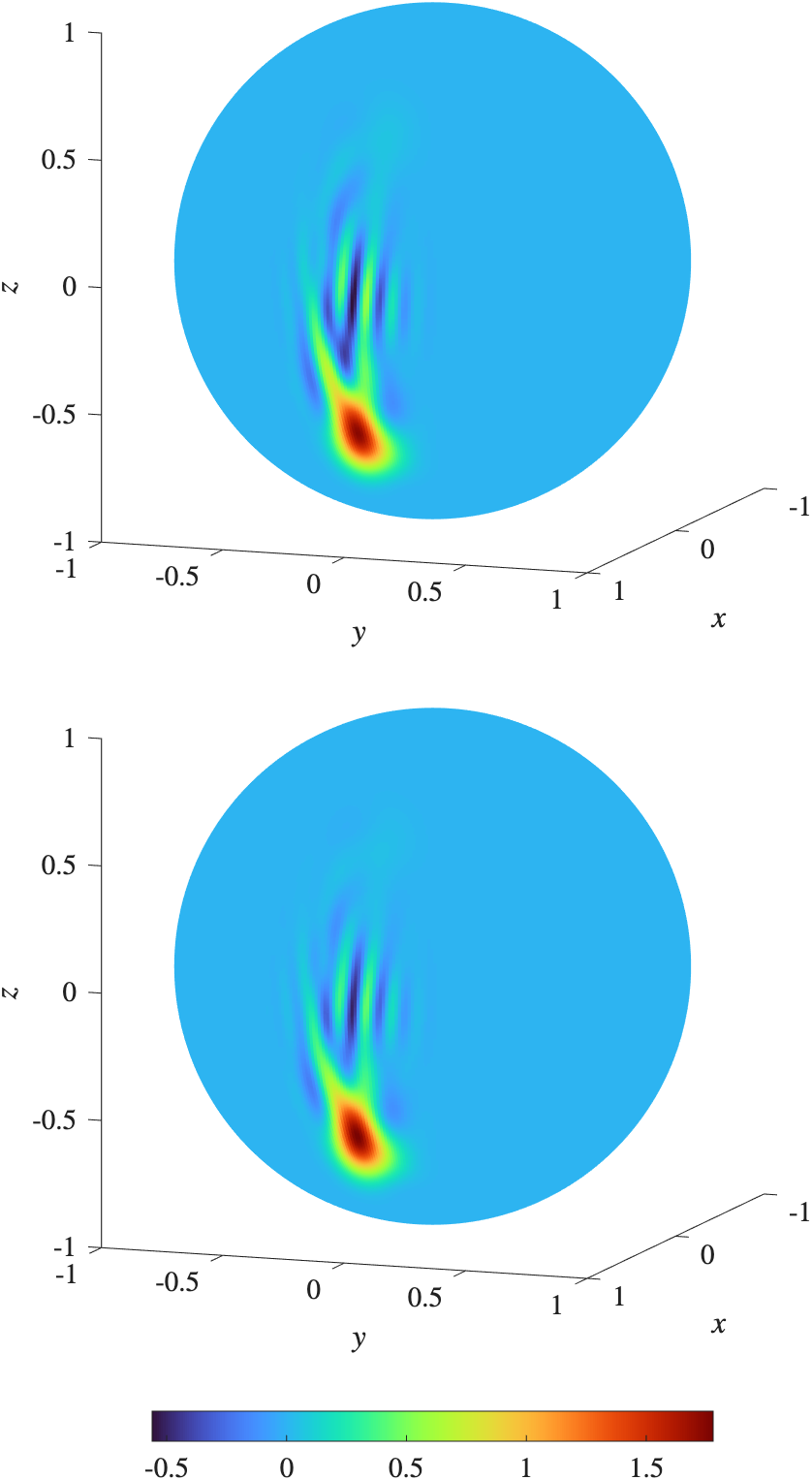}
  
  \caption{Comparison of quantum and TS-IMF propagators  for $j=50$ with a scaling factor of $0.991$ for $\varepsilon=0, \nu=1, \kappa=0.6$ and an initial coherent state at $z_0=0.553$ and $\phi_0=0$. 
The top plot shows the $z-$expectation values, quantum in blue and TS-IMF in black. The bottom panel depicts The Wigner functions of the true (top row) and TS-IMF (bottom row) propagation at  times $t=18, 325, 550, 600$. }
  \label{BHtimesliceWigner}
\end{figure}

 A look at the analytical result for the IMF propagator for the pure interaction term in the limit $j\to\infty$ (\ref{UIMlargejapprox}), 
reveals, that the error in every time-slice is linear in $\delta t$, and in fact accumulates in the time-sliced propagator, and manifests as an effective scaling in the interaction strength. This can be seen as follows. Inserting (\ref{UIMlargejapprox}) into (\ref{TSpropexpression}) we find the matrix elements of the time-sliced IMF propagator for the interaction term in the limit of $j\to\infty$ and the limit of infinitesimal time slices ($\delta t\to 0$) as
\begin{align}
    \hat{U}^{TS}_{n, n'}(t) &= \lim_{M\to \infty} \prod^M_{m=1} \delta_{n n'} e^{- \rmi 2\kappa  n^2  t/ Mj}\left[1 -2 \rmi \kappa\frac{t}{M}\left(1 - \frac{n^2}{j^2}\right)\right]^{-1/2},\nn \\
    &= \delta_{n n'} e^{- 2\rmi \kappa n^2  t/ j} \lim_{M\to \infty}  \left(1 - \frac{a t}{M}\right)^{-M/2},
\end{align}
where $a =2 \rmi \kappa \left(1 - \frac{n^2}{j^2}\right)$. Using that
\begin{equation}
     e^{x} = \lim_{M\to \infty}\left( 1 + \frac{x}{M}\right)^M,   
\end{equation}
this reduces to
\begin{align}
    \hat{U}^{TS}_{n, n'}(t) &= \delta_{n n'} e^{-2\rmi \kappa (1 + \frac{1}{2j}) n^2/j } e^{\rmi \kappa t}.
\end{align}
The second exponential term is a constant real phase and does not affect the state dynamics. The other term has the same form as the exact solution, however, with a rescaled interaction parameter $\kappa_{eff}$ which satisfies
\begin{equation}
\frac{\kappa_{eff}}{\kappa} = \left(1 + \frac{1}{2j}\right)^{-1}=1-\frac{1}{2j+1}
\label{scaling-large-j}
\end{equation}

In figure \ref{BHtimesliceExp_unscaled vs predicted} we show the $z$-expectation value for $j=50$ for $\kappa=0.6$ and an initial coherent state, centred at $z_0=0.553$ and $\phi_0=0$, just shy of the self trapping point. We have chosen a small value of $\kappa$ here, to force some tunnelling behaviour on a reasonable time scale, despite the relatively large particle number. The tunnelling time for the many-particle system is around $600$ time units. Even in the unscaled case, the method clearly picks up on the tunneling, as well as subtle fast oscillations. When the predicted scaling factor of $0.99$ is applied, we see a near reproduction of the `short' time dynamics (up to around $t\approx 80$), and even over long times the agreement is very close.

From numerical testing we have found that the approximation can be further improved with a scaling factor of $0.991$. The resulting expectation value dynamics as well as the Wigner functions for four selected times are depicted in comparison to the exact results in figure \ref{BHtimesliceWigner}, showing an excellent agreement. The Wigner functions produced with the TS-IMF and the adapted scaling factor, in particular, are nearly indistinguishable from the exact ones. 

\begin{figure}[tb]
  \centering
  \includegraphics[width=0.65\textwidth]{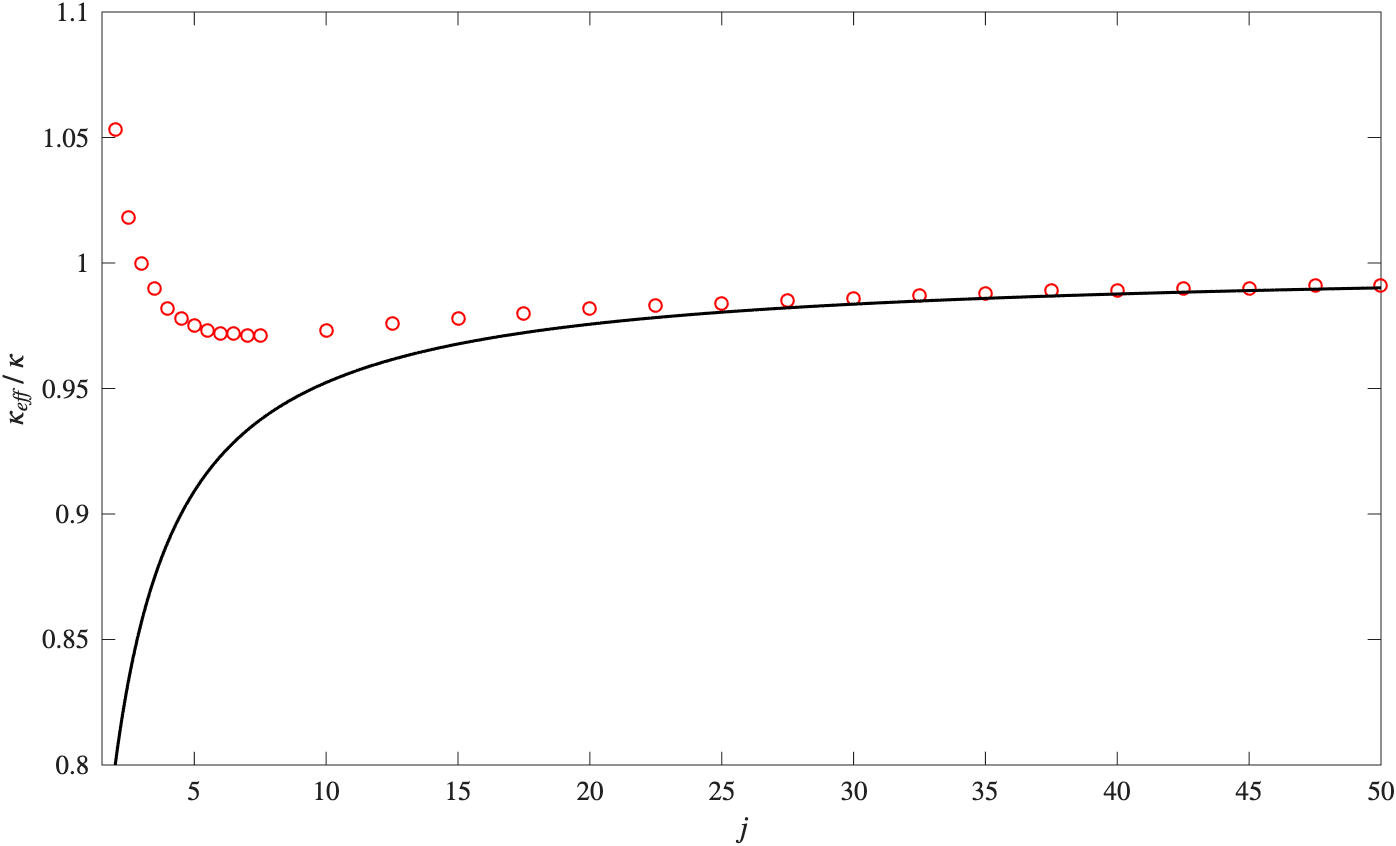}
  \caption{Numerically obtained optimal scaling factor of the interaction strength (red circles) in dependence on $j$, in comparison to the analytical asymptotic behaviour (black line). }
  \label{scaling factor plot}
\end{figure}

Motivated by this observation, we ran a large number of numerical tests on lower values of $j$. From this we found that the TS-IMF propagator does reproduce the exact dynamics to high accuracy up to large times for a rescaled interaction value, where the scaling factor $\kappa_{eff}/\kappa$ does not depend on initial conditions or parameter values, but solely on the value of $j$. Figure \ref{scaling factor plot} shows the numerically obtained optimal scaling factors for a range of values of $j$ in comparison to the large $j$ estimate (\ref{scaling-large-j}). We clearly observe the convergence of the scaling factors to the predicted asymptotic behaviour. 
Further details on the numerical procedure and a comprehensive table are available in \ref{sec-appendix-Scaling}.

\begin{figure}[tb]
  \centering  
  \includegraphics[width=0.65\textwidth]{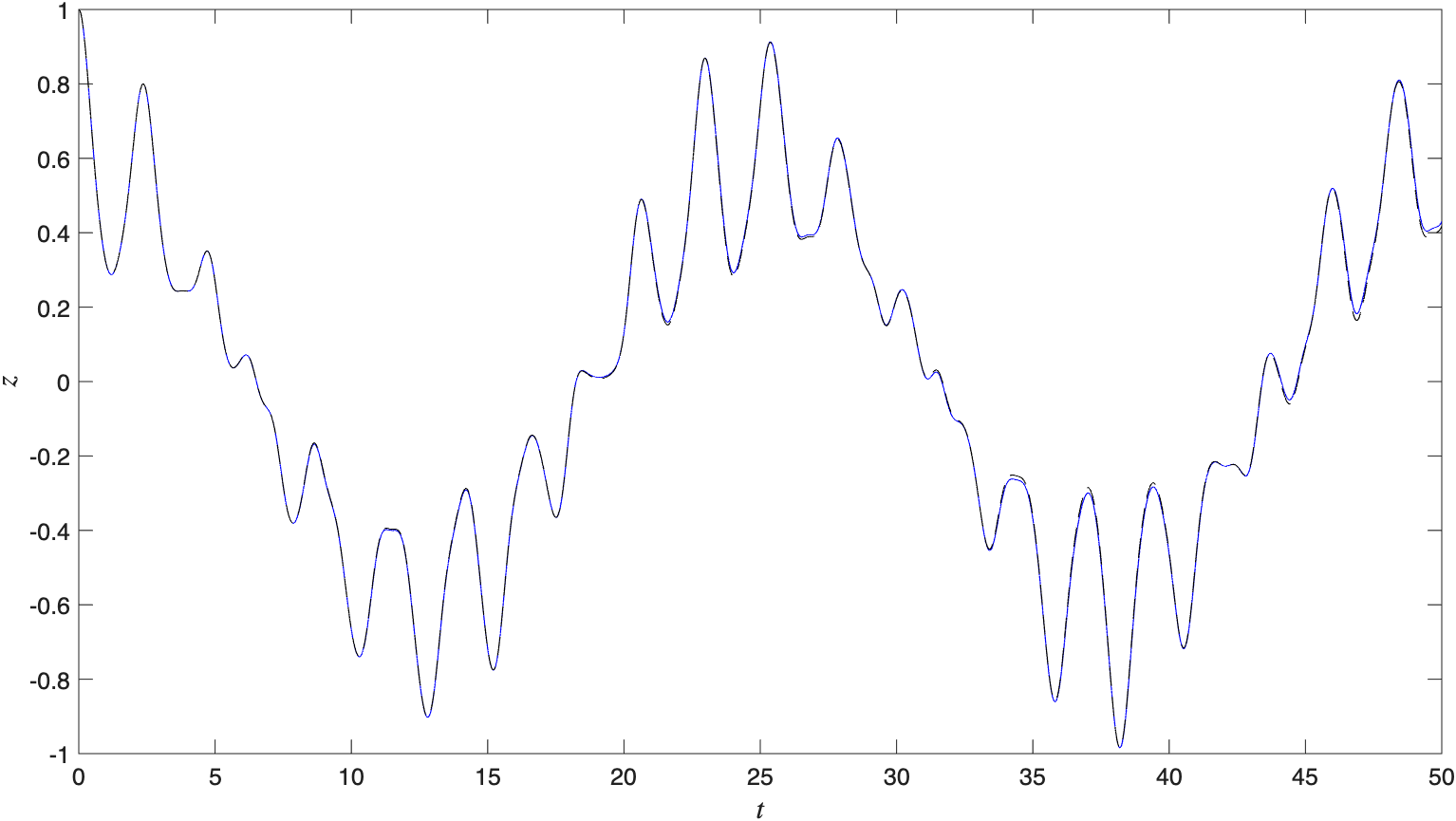}
  
  \vspace{1em}
  
  \includegraphics[width=0.3\textwidth]{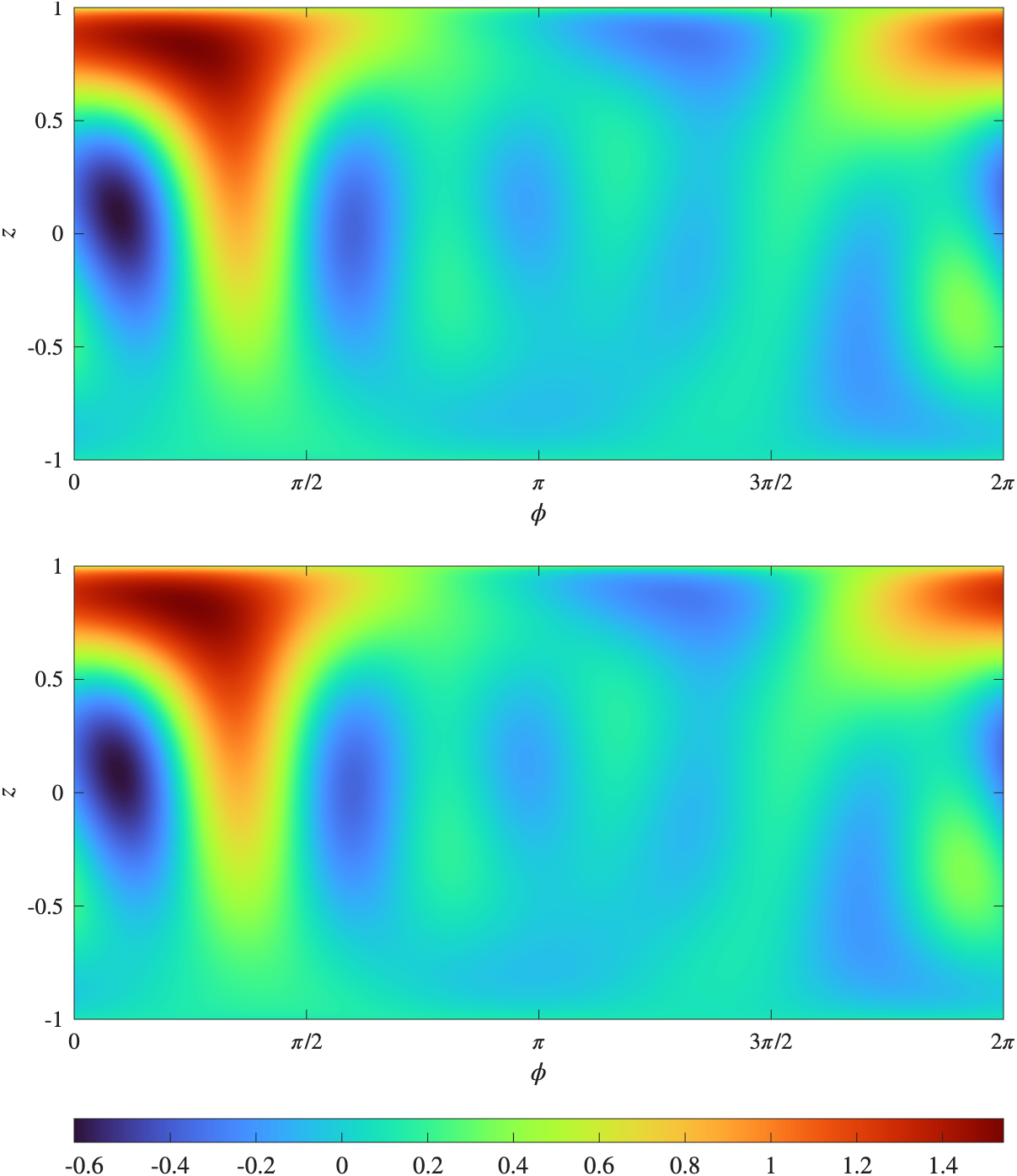}
  \includegraphics[width=0.3\textwidth]{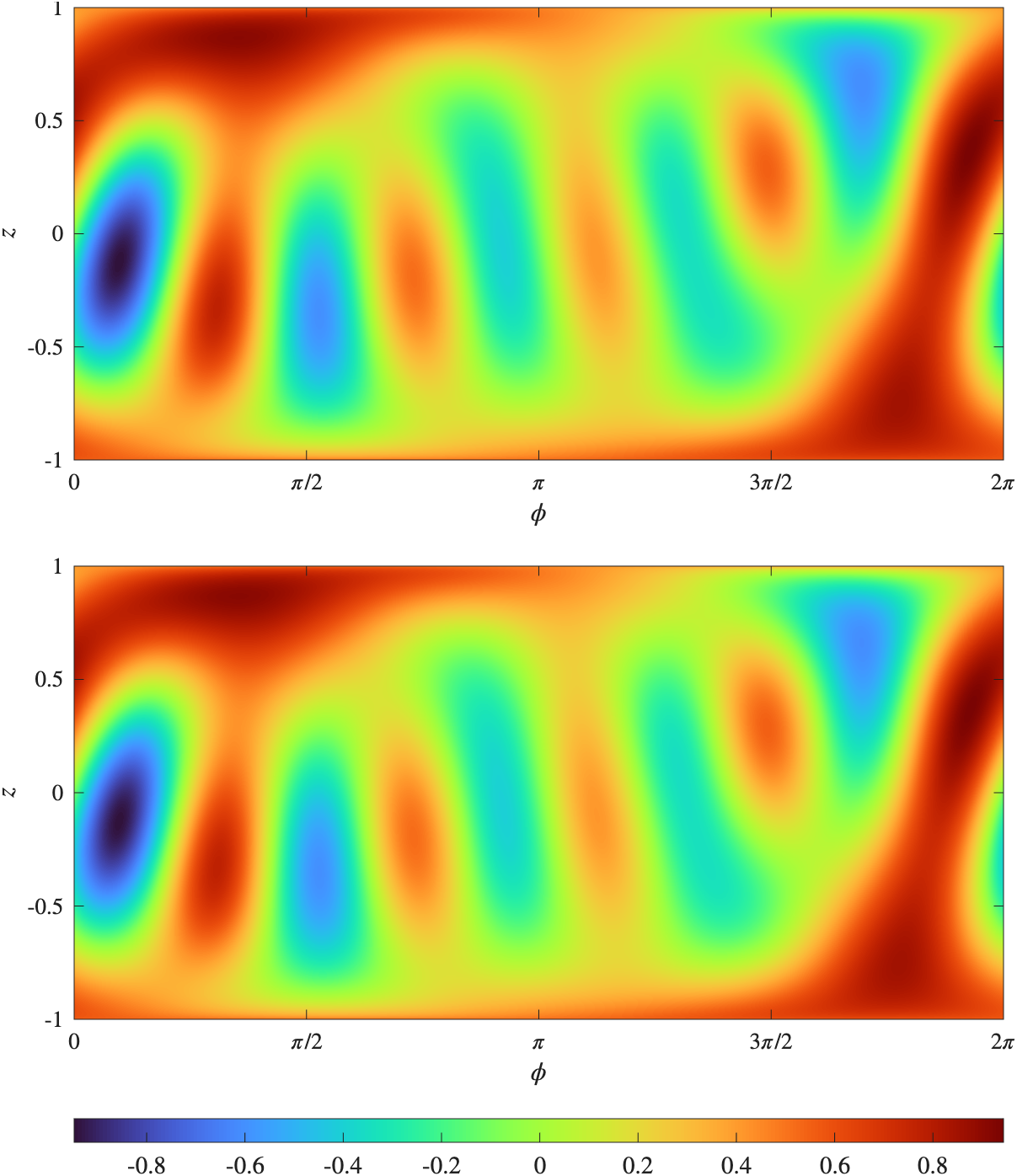}
  \includegraphics[width=0.3\textwidth]{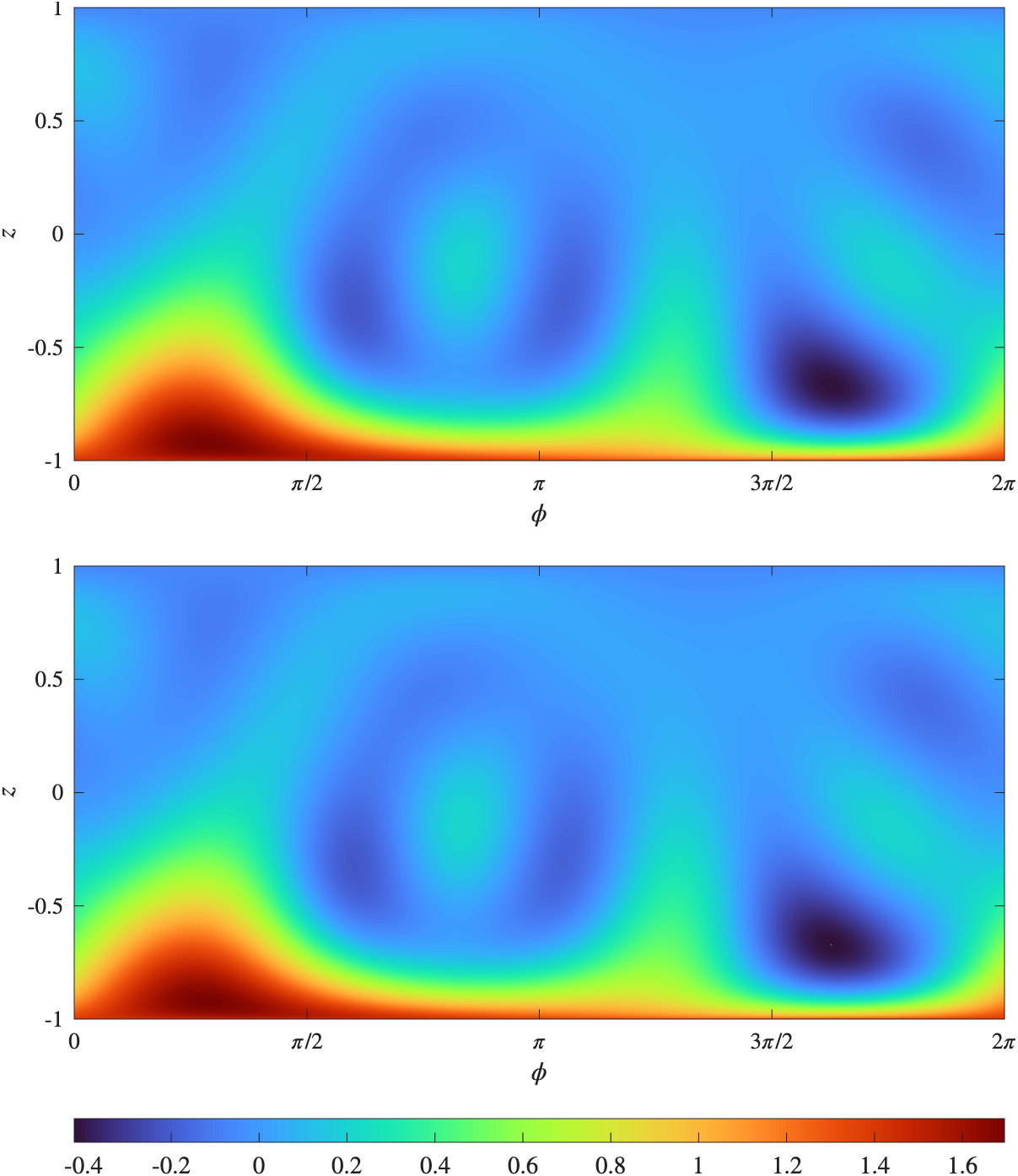}
  \caption{Comparison of quantum and TS-IMF propagators for a spin 2.5 system where a scaling factor of $1.018$ has been used in the TS procedure. Here $\varepsilon=0, \nu=1, \kappa=1.25$ and the initial condition is a coherent state centred at $z_0=1, \phi_0=0$. The top plot shows the $z-$expectation values, quantum in blue and TS-IMF in black. The Wigner functions of the true (top) and TS-IMF (bottom) are shown for times $t=2$ (before tunnelling), $t=6$ (during tunnelling) and $t=13$ (after tunnelling). }
  \label{spin 2.5}
\end{figure}

Using the numerically obtained optimal scaling factor $1.018$, we are able to accurately simulate even a spin $2.5$ system, far from the semiclassical limit. We consider the parameters $\varepsilon=0, \nu=1, \kappa=1.25$ and the initial condition is a coherent state centred at $z_0=1$ and $\phi_0=0$. Figure \ref{spin 2.5} compares the expectation values and Wigner functions of the TS-IMF to the true time evolved state. Here a grid size as small as $10 \times 10$ was sufficient to obtain results that reproduce even the fine details of the full many-body behaviour. The differences in the exact vs the TS-IMF Wigner functions are almost imperceptibly.

\section{Summary and Outlook}
\label{sec-SO}

In summary, we have constructed a relatively simple semiclassical propagator that utilises interfering mean-field trajectories, which we have dubbed the Interfering Mean-Field (IMF) propagator. This method provides an approximation for the full time-evolved quantum state, and is able to effectively capture collapse and revival phenomena. By extending the method with time-slicing, we are able to replicate even many-particle tunnelling to high accuracy. When an additional scaling is applied to the interaction coefficient, the constructed state is quantitatively extremely close to the true time evolution both for large time scales and far from the semiclassical limit.

Though this paper has focused on the application to the Bose-Hubbard dimer, the method straight-forwardly extends to other $\mathrm{SU}(2)$ Hamiltonians. On a theoretical level the extension to $\mathrm{SU}(M)$ systems, describing for example Bose-Hubbard systems with $M$ sites, poses no significant challenge. However, the higher phase-space dimensions make a direct grid method as implemented here less feasible for a larger number of sites. For this case an evaluation of the phase-space integral using Monte-Carlo sampling could be used efficiently \cite{kroninger2023sampling}. The present method is also well suited for generalisations to non-Hermitian Bose-Hubbard systems \cite{graefe2008mean,graefe2010quantum,graefe2013mean}, which will be the topic of a future work.

As mentioned previously, methods such as Herman-Kluk and renormalisation have been utilised in flat space to remedy the loss of unitarity from semiclassical propagation. It could be interesting to explore the translations of these methods to $\mathrm{SU}(2)$ Hilbert space. This may be useful in eliminating the need for a scaling factor observed in the time-slicing method.

\section*{Acknowledgements}
EFTM acknowledges support via a PhD scholarship funded from the EPSRC UDLA 2527 grant (grant number EP/Z534869/1).


\newpage
\appendix

\section{Equivalence of a Condensed State and $\mathrm{SU}(2)$ Coherent State}
\label{firstapp}
We begin with the equation for the an arbitrary condensed state (\ref{consend}) expanded using the binomial formula
\begin{align}
|\psi_1,\psi_2\rangle&=\frac{1}{\sqrt{N!}}\left(\psi_1\hat a_1^\dagger+\psi_2\hat a_2^\dagger\right)^N|0,0\rangle, \\
&=\frac{1}{\sqrt{N!}}\sum^N_{k=0}\frac{N!}{k!(N-k)!}\psi_1^k\psi_2^{N-k}(\hat{a}_1^{\dagger})^k(\hat{a}_2^{\dagger})^{N-k}|0,0\rangle.
\label{consend_app}
\end{align}
We then apply the creation operators to the vacuum state and simplify
\begin{align}
|\psi_1,\psi_2\rangle&= \frac{1}{\sqrt{N!}}\sum^N_{k=0}\frac{N!}{k!(N-k)!}\psi_1^k\psi_2^{N-k}\sqrt{k!(N-k)!}|k,N-k\rangle, \\
&= \sum^N_{k=0}\sqrt{\frac{N!}{k!(N-k)!}}\psi_1^k\psi_2^{N-k}|k,N-k\rangle.
\label{app_2}
\end{align}
Recalling that the state $|k, N-k\rangle$ may be identified with $|j,m\rangle$ where $N=2j$ and $m=\frac{k - (N-k)}{2} = k-j$ we find
\begin{align}
|\psi_1,\psi_2\rangle
&= \sum^{j}_{m=-j}\sqrt{\frac{2j!}{(j+m)!(j-m)!}}\psi_1^{j+m}\psi_2^{j-m}|j,m\rangle.
\label{app_3}
\end{align}
Setting the coefficient to be
\begin{equation}
    \psi_1=\sqrt{\frac{1+z}{2}}, \quad \psi_2=\sqrt{\frac{1-z}{2}}\rme^{\rmi \phi}.
\end{equation}
we see the equivalence of (\ref{consend}) with (\ref{thisone}), namely
\begin{align}
\nn |\zeta\rangle
&=\frac{1}{2^j}\sum_{m=-j}^{j} \sqrt{\binom{2j}{j-m}}\,(1+z)^{(j+m)/2} (1-z)^{(j-m)/2}  e^{\rmi \phi(j-m)}|j,m \rangle.
\end{align}

\section{Expectation Value Derivations}
\label{sec-CS-expval}

To begin we derive the action of the spin raising and lowering operators $\hat{J}_\pm = \hat{J}_x \pm \rmi \hat{J}_y$ as well as $\hat{J}_z$ on an unnormalised coherent state $||\zeta\rangle = e^{\zeta\hat{J}_-} |j, j\rangle$. The Hadamard lemmas will be useful for this exercise
\begin{align}
    e^{-\zeta \hat{J}_-} \hat{J}_ze^{\zeta \hat{J}_-} &=\hat{J}_z - \zeta \hat{J}_-, \label{Had1}\\
    e^{-\zeta \hat{J}_+} \hat{J}_- e^{\zeta \hat{J}_+} &= \hat{J}_- - 2 \zeta \hat{J}_z - \zeta^2 \hat{J}_+, \label{Had2}\\
    e^{-\zeta \hat{J}_-} \hat{J}_+ e^{\zeta \hat{J}_-} &= \hat{J}_+ + 2 \zeta \hat{J}_z - \zeta^2 \hat{J}_-.\label{Had3}
\end{align}

Finding the action of the lowering operator on an unnormalised coherent state is straightforward:
\begin{align}
    \hat{J}_-||\zeta\rangle &= \hat{J}_-e^{\zeta\hat{J}_-}|j, j\rangle , \nn\\
    &= \frac{\partial}{\partial\zeta}\left\{e^{\zeta\hat{J}_-}\right\}|j, j\rangle, \nn\\
    &= \frac{\partial}{\partial\zeta}||\zeta\rangle.\label{Jminusaction}
\end{align}
To find the action of the raising operator, we employ (\ref{Had3})
\begin{align}
    \hat{J}_+||\zeta\rangle &=\hat{J}_+e^{\zeta\hat{J}_-}|j, j\rangle, \nn\\
    &= e^{\zeta \hat{J}_-}\left(\hat{J}_+ + 2\zeta \hat{J}_z - \zeta^2\hat{J}_-  \right)|j, j\rangle, \nn \\
    &= \left(2\zeta j - \zeta^2\frac{\partial}{\partial\zeta}  \right)||\zeta\rangle. \label{Jplusaction}
\end{align}
In a similar fashion we employ (\ref{Had1}) to find the action of $\hat{J}_z$, 
\begin{align}
    \hat{J}_z||\zeta\rangle &=\hat{J}_ze^{\zeta\hat{J}_-}|j, j\rangle, \nn\\
    &= e^{\zeta \hat{J}_-}\left(\hat{J}_z - \zeta \hat{J}_-   \right)|j, j\rangle, \nn \\
    &= \left(j - \zeta  \frac{\partial}{\partial\zeta} \right) ||\zeta\rangle.
\end{align}

With the above the expectation values of these three operators in normalised coherent states may be found. Defining the function 
\begin{align}
    M(\zeta, \zeta^*) = \langle\zeta|| ||\zeta\rangle = (1 + \zeta\zeta^*)^{2j}, 
\end{align}
the expectation value of $\hat{J}_-$ may be expressed as
\begin{align}
    \langle \hat{J}_- \rangle &= \frac{\langle \zeta|| \hat{J}_-||\zeta\rangle}{M(\zeta, \zeta^*)}, \nn \\
    &= \frac{\langle \zeta|| \frac{\partial}{\partial \zeta}||\zeta\rangle}{M(\zeta, \zeta^*)}, \nn \\
    &=\frac{1}{M}\frac{\partial M}{\partial \zeta} \nn, \\
    &=\frac{2 j \zeta^* }{1 + | \zeta|^2}, \label{Jminusexp_app}
\end{align}
where we may move the partial derivative to the left of the bra state as this is only a function of $\zeta^*$. With similar reasoning, or by taking the complex conjugate, we find the expectation of the raising operator
\begin{align}
    \langle \hat{J}_+ \rangle 
    &=\frac{2 j \zeta }{1 + | \zeta|^2}. \label{Jpluexp}
\end{align}
The expectation value of $\hat{J}_z$ is found as
\begin{align}
    \langle \hat{J}_z \rangle 
    &=j\frac{1 - |\zeta|^2}{1 + | \zeta|^2}.
\end{align}

Noting  that
\begin{align}
    \hat{J}_x = \frac{1}{2}\left(\hat{J}_+ + \hat{J}_-\right), \quad \hat{J}_y = \frac{1}{2\rmi}\left(\hat{J}_+ - \hat{J}_-\right),
\end{align}
from (\ref{Jminusaction}) and (\ref{Jplusaction}) the action of $\hat{J}_x$ and $\hat{J}_y$ on an unnormalised coherent state is found to be
\begin{align}
    \hat{J}_x || \zeta \rangle &= \frac{1}{2}\left(2 \zeta j + (1 - \zeta^2)\frac{\partial}{\partial \zeta}\right)|| \zeta \rangle, \\
    \hat{J}_y || \zeta \rangle &= \frac{1}{2\rmi}\left(2 \zeta j - (1 + \zeta^2)\frac{\partial}{\partial \zeta}\right)||\zeta \rangle, 
\end{align}
and from (\ref{Jminusexp_app}) and (\ref{Jpluexp}) their expectation values in coherent states may be found as
\begin{align}
    \langle \hat{J}_x \rangle &= \frac{j(\zeta + \zeta^*)}{1+|\zeta|^2}, \\
    \langle \hat{J}_y \rangle &= \frac{j(\zeta - \zeta^*)}{\rmi(1+|\zeta|^2)}.
\end{align}

For an arbitrary operator $\hat{A}$, we may use the action of the spin operators to find the coherent state expectation values of $A\hat{J}_k$ and $\hat{J}_kA$, where $k=x, y, z, +, -$. These can be concisely summarised as
\begin{align}
    \langle   \hat{A} \hat{J}_k  \rangle &= \langle  \hat{J}_k  \rangle\langle   \hat{A}  \rangle + \frac{(1 + |\zeta|^2)^2}{2j}\frac{\partial \langle    \hat{J}_k  \rangle}{\partial \zeta}\frac{\partial \langle    \hat{A}  \rangle}{\partial \zeta^*}, \\
    \langle    \hat{J}_k  \hat{A} \rangle &= \langle  \hat{J}_k  \rangle\langle   \hat{A}  \rangle + \frac{(1 + |\zeta|^2)^2}{2j}\frac{\partial \langle    \hat{J}_k  \rangle}{\partial \zeta^*}\frac{\partial \langle    \hat{A}  \rangle}{\partial \zeta}
    \label{expJkA}
\end{align}

\section{Derivation of the SCS-TDVP Equations}
\label{sec-appendix-TDVP}
In this appendix we provide a derivation for the equations of motions for the parameters in the SCS-TDVP aaproximation. 

We consider an initially coherent state $|\psi_0\rangle = c_0 | \zeta_0\rangle$, 
and make the ansatz
\begin{align}
    | \psi(t) \rangle = c(t) | \zeta(t)\rangle \text{ such that } | \psi(0) \rangle = c_0 | \zeta_0\rangle \label{constraint-app}
\end{align}
for the solution  of the time-dependent Schr\"odinger equation.

The dynamical equations for the parameters can be derived via the Dirac-Frenkel variational principle \cite{kramer2008review,shalashilin2008gaussian,werther2021coherent}, by demanding that
\begin{align}
    \rmi \langle \delta \psi |   \dot{\psi} \rangle =  \langle \delta \psi | \hat{H}| \psi \rangle, \label{DFV}
\end{align}
where $|\psi(t)\rangle$ is of the from (\ref{constraint-app}), and the variation $\langle \delta \psi|$ is given by
\begin{equation}
  \langle  \delta \psi | = \delta c^*\frac{\partial \langle \psi |}{\partial c^*} +\delta \zeta \frac{\langle \psi |}{\partial \zeta} + \delta \zeta^* \frac{\langle \psi |}{\partial \zeta^*}.
\end{equation}
Using the coherent-state equalities
\begin{align}
    \frac{\partial \langle \zeta|}{\partial \zeta} &= -\frac{j \zeta^*}{1 + |\zeta|^2} \langle \zeta|\\
    \frac{\partial \langle \zeta|}{\partial \zeta^*} &= \langle \zeta| \Big[-  \frac{ j \zeta}{1 + |\zeta|^2} + \hat{J}_+ \Big] 
\end{align}
this can be expressed as 
\begin{align}
\label{variation}
    \langle  \delta \psi | &= \langle \zeta | \Big[\delta c^* -c^* \delta \zeta \frac{j \zeta^*}{1 + |\zeta|^2} + c^* \delta \zeta^*\Big(-  \frac{ j \zeta}{1 + |\zeta|^2} + \hat{J}_+ \Big) \Big].
\end{align}
Inserting (\ref{variation}) into (\ref{DFV}) yields a linear equation for the variations of $c, \zeta$ and $\zeta^*$. Since these variations are mutually independent \cite{werther2021coherent}, we may equate coefficients. The $\delta c$ and $\delta \zeta$ terms yield the same equation 
\begin{align}
\label{deltac-equation}
    \rmi \langle \zeta |  \dot{\psi}\rangle  &= \langle \zeta | \hat{H} | \psi \rangle,
\end{align}
while the $\delta \zeta^*$ terms leads to
\begin{align}
    \rmi  c^* \langle \zeta | \hat{J}_+ | \dot{\psi} \rangle &= c^* \langle \zeta |\hat{J}_+ \hat{H}|\psi \rangle.
\end{align}
The time derivative of the ansatz state $|\psi\rangle$ in terms of the parameters $c$, $\zeta$, and $\zeta^*$ is given by
\begin{equation}
\label{dotvariation}
    | \dot{\psi} \rangle  =   \dot{c}| \zeta \rangle + c\dot{\zeta} \frac{\partial | \zeta \rangle}{\partial \zeta} + c\dot{\zeta}^* \frac{\partial | \zeta \rangle}{\partial \zeta^*}, 
\end{equation}
which, using
\begin{align}
    \frac{\partial | \zeta\rangle}{\partial \zeta^*} &= -\frac{j \zeta}{1 + |\zeta|^2}  |\zeta\rangle\\
    \frac{\partial | \zeta\rangle}{\partial \zeta} &=  \Big[-  \frac{ j \zeta^*}{1 + |\zeta|^2} + \hat{J}_- \Big] |\zeta\rangle,
\end{align}
becomes
\begin{equation}
   | \dot{\psi} \rangle =  \Big[ \dot{c} - \frac{ jc }{1 + |\zeta|^2} \Big( \zeta^* \dot{\zeta} + \zeta \dot{\zeta}^*\Big) + c\dot{\zeta}\hat{J}_-  \Big] | \zeta \rangle.\label{timeDervPsi}
\end{equation}
Inserting this into the $\delta c$ coefficient equation (\ref{deltac-equation}), some algebra yields
\begin{equation}
    \rmi  \dot{c} = c \langle \hat H\rangle + \rmi  \frac{c j }{1 + | \zeta|^2}\Big[ \zeta \dot{\zeta}^* - \zeta^* \dot{\zeta}\Big], \label{cdotequation}
\end{equation}
 where we have used the abbreviation $\langle \hat H\rangle=\langle \zeta|\hat H|\zeta\rangle$ to denote the expectation value of the Hamiltonian in coherent states. 
 
Turning to the $\delta \zeta$ coefficient equation, by inserting (\ref{timeDervPsi}) we find
\begin{equation}
    \rmi  c^* \Big[ \dot{c} -  \frac{j c}{1 + |\zeta|^2}\Big(\zeta\dot{\zeta}^* + \zeta \dot{\zeta}^* \Big) \Big] \langle \hat{J}_+ \rangle + \rmi  c c^* \dot{\zeta} \langle \hat{J}_+\hat{J}_- \rangle = cc^* \langle \hat{J}_+ \hat{H}\rangle. \label{deltamucoeff}
\end{equation}
With (\ref{expJkA}), (\ref{deltamucoeff}) reduces to
\begin{equation}
    \rmi \dot{c}  \frac{2j \zeta c^*}{1+|\zeta|^2 } +\rmi  \frac{2j c c^*}{[1 + |\zeta|^2]^2}\big[j \zeta(\zeta^* \dot{\zeta} - \zeta \dot{\zeta}^*)  + \dot{\zeta}\big] =  c c^*\Big[ \frac{2j \zeta}{1 + |\zeta|^2}\langle \hat{H}\rangle + \frac{\partial \langle \hat{h}\rangle}{\partial \zeta^*} \Big].
\end{equation}
Inserting (\ref{cdotequation}) and rearranging, we find
\begin{align}
    \rmi  c c^* \frac{2j}{[1 + |\zeta|^2]^2}\dot{\zeta} &= c c^* \frac{\partial \langle \hat{H}\rangle}{\partial \zeta^*}.
\end{align}
As $|c|^2 \neq 0 $, we arrive at the equations of motion for the prefactor and coherent state co-ordinate, 
\begin{align}
    \rmi  \dot{\zeta} &= \frac{(1 + |\zeta|^2)^2}{2j} \frac{\partial \langle \hat{H}\rangle}{\partial \zeta^*}, \label{muEOM-app}\\
    \rmi  \dot{c} &= c \Big[ \langle \hat{H}\rangle + \rmi  \frac{j}{1 + |\zeta|^2}( \zeta^* \dot{\zeta} - \zeta \dot{\zeta}^*)\Big] \label{cEOMwithzeta-app}
\end{align} 

\section{Saddle point approximation for the IMF propagator for pure interaction}
\label{app-saddle}

The integrand of $g_m(A)$ (equation (\ref{equ-gma})) is highly oscillatory, motivating the use of the saddle point approximation \cite{bender1999advanced}. To begin we assume $m\neq \pm j$. Recalling that $\beta$ does not scale with $j$, we may write $g_m(A)$ in the form
\begin{align}
    g_m(A) = \frac{2j+1}{2^{2j+1}} \binom{2j}{j-m}\int^1_{-1}  e^{j \Phi(z_0)}\rmd z_0,
\end{align}
where
\begin{align}
    \Phi(z_0) = (1 + B)\ln(1 + z_0) + (1 - B)\ln(1- z_0)  + \rmi A(z_0 - B)^2.
\end{align}
and we have defined $B = m/j \in (-1,1)$, which we assume to be fixed. The dominant contributions of the integrand come from the maxima of $\Phi(z_0)$  for real $z_0$ which lie within the integration bounds. We find those by evaluating the derivatives
\begin{align}
    \Phi'(z_0) &= \frac{1 + B}{1 +z_0} - \frac{1 - B}{1 - z_0} + 2 \rmi A(z_0-B), \nn \\
    \Phi''(z_0) &= - \frac{1 + B}{(1+z_0)^2}-\frac{1-B}{(1-z_0)^2} + 2\rmi A. \nn
\end{align}
From these we find the real saddle to be $z^*=B$. The general form of the approximation will be \cite{bender1999advanced}
\begin{align}
    g_m(A) = \frac{2j+1}{2^{2j+1}} \binom{2j}{j-m}\sum_{\text{saddle points } z^*} e^{j \Phi(z^*)}\sqrt{\frac{2 \pi}{- j \Phi''(z^*)}}\Big[1  + O(1/j)\Big].
\end{align}
Evaluating the first and second derivatives at the saddle point and inserting them into this formula yields
\begin{align}
    g_m(A) =& \frac{2j+1}{2^{2j+1}} \binom{2j}{j-m}e^{j (1 +B))\ln(1+B) + j(1-B)\ln(1-B)} \nn \\
    \times& \sqrt{\frac{\pi(1+B)(1-B)}{j[1 -  \rmi A(1-B^2)]}}\Big[1  + O(1/j)\Big].
\end{align}
For large $j$, Stirling's approximation may be used to show
\begin{align}
    \frac{2j+1}{2^{2j+1}}  \binom{2j}{j-m} \approx \frac{j}{2^{2j}}\cdot\frac{e^{-j[(1-B)\ln(1-B) + (1 + B)\ln(1+B) - 2\ln2]}}{\sqrt{\pi j (1-B)(1+B)}} 
\end{align}
which leads to our function becoming approximately 
\begin{align}
    g_m(A) &\approx \frac{j}{2^{2j}} \cdot \frac{e^{2j\ln2}}{j\sqrt{1 - \rmi A(1-B^2)}} \Big[1  + O(1/j)\Big], \nn\\
    &= \frac{1}{\sqrt{1 - \rmi A(1-B^2)}}+ O(1/j). 
\end{align}
Recalling that IMF propagator matrix elements are 
\begin{align}
    U_{n,m}^{IMF} = \delta_{nm}e^{-2\rmi  \kappa t m^2 / j} g_m(2t \kappa),
\end{align}
in the large $j$ limit this is approximated by
\begin{align}
    U^{IMF}_{n,m}(t; \text{large }j) &=  \frac{\delta_{nm} e^{-2\rmi  \kappa t m^2 / j}}{\sqrt{1 -2\rmi t \kappa(1-\frac{m^2}{j^2})}}. \label{UIMlargejapprox-app} 
\end{align}

In the case of $m=j$, $B=1$ and the function reduces to 
\begin{align}
    g_j(A) = \frac{2j+1}{2^{2j+1}} \int^1_{-1}  e^{j f(z_0)}\rmd z_0,
\end{align}
with
\begin{align}
    f(z_0) = 2\ln(1 + z_0) +  \rmi A(z_0 - 1)^2.
\end{align}
In this case there is no saddle on the real axis. The integrand vanishes at the lower integral limit, $z_0=-1$, and it's real part is maximal at the upper integration limit, $z_0=1$. Thus, the integral can be approximated by a Laplace method \cite{bender1999advanced}, where the leading contribution comes from the upper limit of the integral, and we find
\begin{align}
    g_j(A) &\approx \frac{j}{2^{2j}} \frac{e^{j f(1)}}{j f'(1)}, \nn \\
    &= 1.
\end{align}
The case where $m=-j$ follows in a similar fashion and yields the same result. Thus the $m=\pm j$ elements follow the formula (\ref{UIMlargejapprox-app}) and are identical to the true matrix elements.

\section{Wigner Functions on the Sphere}
\label{sec-appendix-wigner}

To visualise the quantum state $|\psi\rangle$ we use the $\mathrm{SU}(2)$ Wigner function defined as
\cite{klimov2017generalized} 
\begin{align}
W(\theta, \phi) = {\rm tr}\left(|\psi\rangle\langle \psi| \, \hat{w}(\theta,\phi)\right) 
\label{eq:6} 
\end{align}
on the spherical phase space, where the Stratonovich-Weyl kernel is given by
\begin{align}
\hat{w}(\theta, \phi) = \frac{1}{\sqrt{2j+1}} 
\sum_{l=0}^{2j} \sum_{k=-l}^{l}
Y^k_{l}(\theta, \phi) \, {\hat T}^{\,j}_{l,k} \nonumber
\label{eq:7}
\end{align}
Here $\{\hat T^{\,j}_{l,k}\}$ are the 
irreducible tensor operators
\cite{edmonds1996angular}, and 
$Y^{k}_{l}(\theta, \phi)$ denote the spherical harmonics
\cite{edmonds1996angular}.

The matrix elements of the irreducible 
tensors in the standard $\hat J_z$-basis are given by
\begin{align}
\langle j,m'|\hat T^{j}_{l,k}|j,m\rangle=\sqrt{\frac{2l+1}{2j+1}}\, 
C_{jmlk}^{jm'} \, ,
\end{align}
where the $C_{j_1m_1j_2m_2}^{jm}$ denote the Clebsch-Gordan coefficients. 
 
We use the convention $Y^0_0=1$, 
so that the spherical harmonics are orthonormal with respect to the uniform 
probability measure $\rmd\zeta^0_{\theta,\phi}=(4\pi)^{-1}\sin\theta\,\rmd\theta\,\rmd\phi$, that is, it holds
\begin{align}
\int Y^{k_1}_{L_1} \overline{Y^{k_2}_{L_2}}\,\rmd \zeta^0_{\theta,\phi}
= \delta_{l_1l_2}\,\delta_{k_1 k_2} \, .
\label{eq:x8}
\end{align}

\section{Scaling Factors for the Non-Linear Coefficient}
\label{sec-appendix-Scaling}

\begin{table}[tb]
\begin{center}
\begin{tabular}{|c | c | c || c|c|c||c|c|c|} 
 \hline
 $j$ & $\kappa_{eff} / \kappa$ & $(1+\frac{1}{2j})^{-1}$ & $j$ & $\kappa_{eff} / \kappa$ & $(1+\frac{1}{2j})^{-1}$ & $j$ & $\kappa_{eff} / \kappa$ & $(1+\frac{1}{2j})^{-1}$ \\
 \hline
 2   & 1.053 & 0.800 & 7.5  & 0.971 & 0.937 & 25   & 0.984 & 0.980 \\
 2.5 & 1.018 & 0.833 & 8    & 0.972 & 0.941 & 27.5 & 0.985 & 0.982 \\
 3   & 1.000 & 0.857 & 8.5  & 0.972 & 0.944 & 30   & 0.986 & 0.983 \\
 3.5 & 0.990 & 0.875 & 9    & 0.972 & 0.947 & 32.5 & 0.987 & 0.985 \\
 4   & 0.982 & 0.889 & 9.5  & 0.973 & 0.950 & 35   & 0.988 & 0.986 \\
 4.5 & 0.978 & 0.900 & 10   & 0.973 & 0.952 & 37.5 & 0.989 & 0.987 \\
 5   & 0.975 & 0.909 & 12.5 & 0.976 & 0.962 & 40   & 0.989 & 0.987 \\
 5.5 & 0.973 & 0.917 & 15   & 0.978 & 0.968 & 42.5 & 0.990 & 0.988 \\
 6   & 0.972 & 0.923 & 17.5 & 0.980 & 0.972 & 45   & 0.990 & 0.989 \\ 
 6.5 & 0.972 & 0.929 & 20   & 0.982 & 0.976 & 47.5 & 0.991 & 0.989 \\
 7   & 0.971 & 0.934 & 22.5 & 0.983 & 0.978 & 50   & 0.991 & 0.990 \\
 \hline
\end{tabular}
\end{center}
\caption{\label{table!!} Numerically obtained optimal scaling factors versus the large $j$ prediction for various spin values.}
\end{table}
\vspace{1em}

We numerically deduced the scaling factors for the interaction coefficients for the time-sliced propagator summarised in table \ref{table!!}. To deduce these, for each value of $j$, we picked various initial conditions and a range of parameter values, some which we know result in tunnelling dynamics and some which were randomly initialised. We ran the time slicing code once and then used a least-square optimisation of the $x, y$ and $z$ expectation values over time to find the optimal scaling value to three decimal places.

\begin{figure}[tb]
  \centering
  \includegraphics[width=0.41\textwidth]{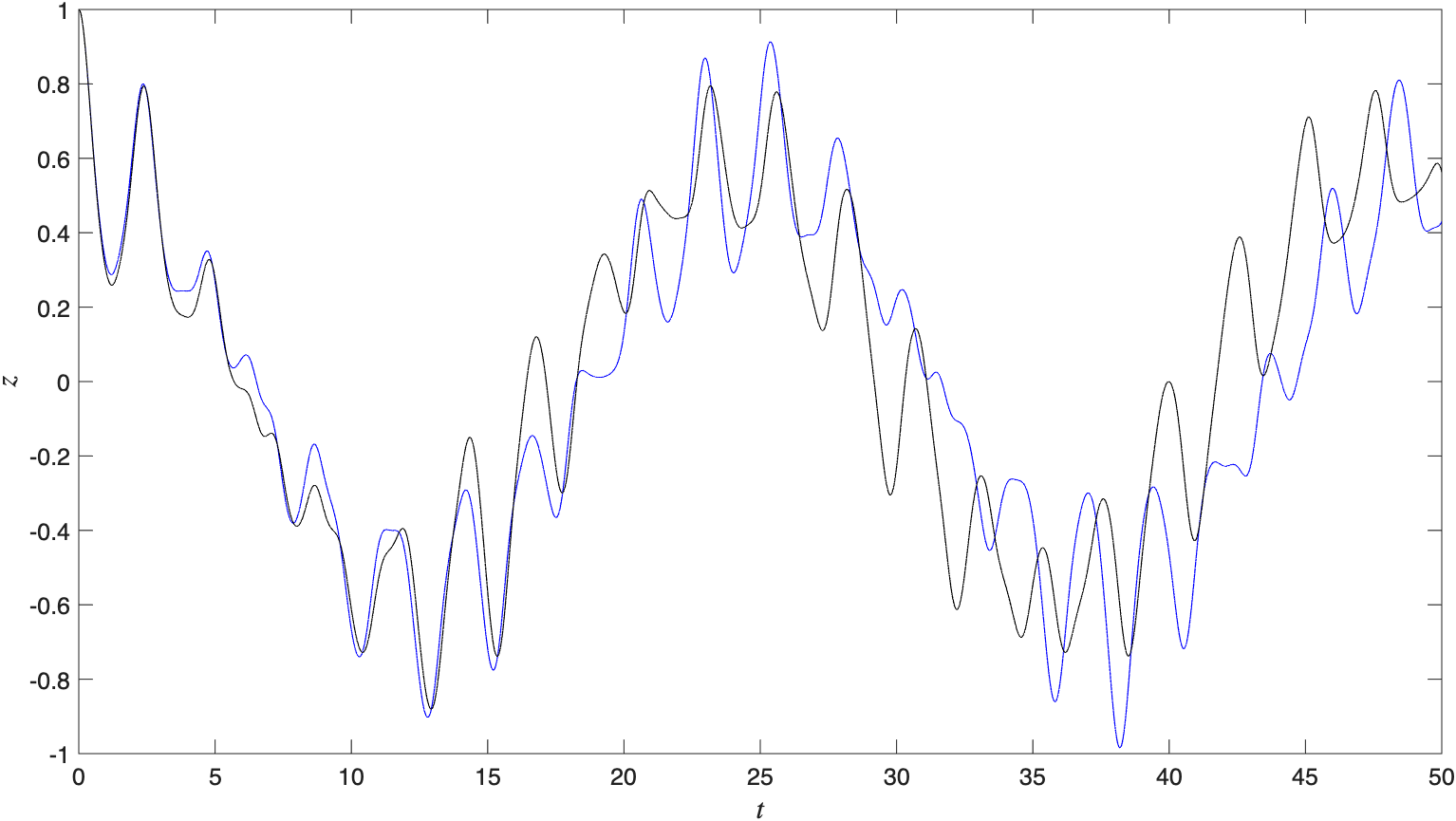}
  \hspace{0.5em}
  \includegraphics[width=0.41\textwidth]{fig11a.png}
  \vspace{1em}
  \includegraphics[width=0.4\textwidth]{fig8a.png}
  \hspace{0.5em}
  \includegraphics[width=0.42\textwidth]{fig9a.png}
  \caption{Comparison of quantum (blue) and TS (black) expectation values for a spin 2.5 (top) and spin 50 (bottom) systems with parameters given in figures \ref{spin 2.5} and \ref{BHtimesliceExp_unscaled vs predicted} respectively. The left column uses an unscaled $\kappa$ in the TS dynamics, while the right column uses the numerically found ideal scaling factor.}
  \label{app-exp}
\end{figure}

In figure \ref{app-exp} we show a direct comparison between the unscaled and scaled expectation values of the two examples discussed in section \ref{sec-TS}.
\providecommand{\newblock}{}


\begin{thebibliography}{10}
\expandafter\ifx\csname url\endcsname\relax
  \def\url#1{{\tt #1}}\fi
\expandafter\ifx\csname urlprefix\endcsname\relax\def\urlprefix{URL }\fi
\providecommand{\eprint}[2][arXiv]{#1:\linebreak[0]#2}

\bibitem{fisher1989boson}
Fisher M~P, Weichman P~B, Grinstein G and Fisher D~S 1989 Boson localization and the superfluid-insulator transition {\em Physical Review B\/} {\bf 40} 546

\bibitem{greiner2002collapse}
Greiner M, Mandel O, H{\"a}nsch T~W and Bloch I 2002 Collapse and revival of the matter wave field of a {B}ose-{E}instein condensate {\em Nature\/} {\bf 419} 51

\bibitem{greiner2002quantum}
Greiner M, Mandel O, Esslinger T, H{\"a}nsch T~W and Bloch I 2002 Quantum phase transition from a superfluid to a {M}ott insulator in a gas of ultracold atoms {\em Nature\/} {\bf 415} 39

\bibitem{bloch2005ultracold}
Bloch I 2005 Ultracold quantum gases in optical lattices {\em Nature Physics\/} {\bf 1} 23

\bibitem{jaksch2005cold}
Jaksch D and Zoller P 2005 The cold atom {H}ubbard toolbox {\em Annals of Physics\/} {\bf 315} 52

\bibitem{polkovnikov2011colloquium}
Polkovnikov A, Sengupta K, Silva A and Vengalattore M 2011 Colloquium: {N}onequilibrium dynamics of closed interacting quantum systems {\em Reviews of Modern Physics\/} {\bf 83} 863

\bibitem{engl2014coherent}
Engl T, Dujardin J, Arg{\"u}elles A, Schlagheck P, Richter K and Urbina J~D 2014 Coherent backscattering in {F}ock space: a signature of quantum many-body interference in interacting bosonic systems {\em Physical Review Letters\/} {\bf 112} 140403

\bibitem{kolovsky2016bose}
Kolovsky A~R 2016 Bose-{H}ubbard {H}amiltonian: quantum chaos approach {\em International Journal of Modern Physics B\/} {\bf 30} 1630009

\bibitem{tomsovic2018post}
Tomsovic S, Schlagheck P, Ullmo D, Urbina J~D and Richter K 2018 Post-{E}hrenfest many-body quantum interferences in ultracold atoms far out of equilibrium {\em Physical Review A\/} {\bf 97} 061606

\bibitem{richter2022semiclassical}
Richter K, Diego~Urbina J and Tomsovic S 2022 Semiclassical roots of universality in many-body quantum chaos {\em Journal of Physics A\/} {\bf 55} 453001

\bibitem{gross2017quantum}
Gross C and Bloch I 2017 Quantum simulations with ultracold atoms in optical lattices {\em Science\/} {\bf 357} 995

\bibitem{daley2022practical}
Daley A~J, Bloch I, Kokail C, Flannigan S, Pearson N, Troyer M and Zoller P 2022 Practical quantum advantage in quantum simulation {\em Nature\/} {\bf 607} 667

\bibitem{eilbeck1985discrete}
Eilbeck J~C, Lomdahl P and Scott A~C 1985 The discrete self-trapping equation {\em Physica D: Nonlinear Phenomena\/} {\bf 16} 318--338

\bibitem{scott1986quantized}
Scott A~C and Eilbeck J 1986 The quantized discrete self-trapping equation {\em Physics Letters A\/} {\bf 119} 60--64

\bibitem{milburn1997quantum}
Milburn G, Corney J, Wright E~M and Walls D 1997 Quantum dynamics of an atomic {B}ose-{E}instein condensate in a double-well potential {\em Physical Review A\/} {\bf 55} 4318

\bibitem{albiez2005direct}
Albiez M, Gati R, F{\"o}lling J, Hunsmann S, Cristiani M and Oberthaler M~K 2005 {D}irect {O}bservation of {T}unneling and {N}onlinear {Se}lf-{T}rapping in a {S}ingle {B}osonic {J}osephson {J}unction {\em Physical Review Letters\/} {\bf 95} 010402

\bibitem{gati2007bosonic}
Gati R and Oberthaler M~K 2007 A bosonic {J}osephson junction {\em Journal of Physics B\/} {\bf 40} R61

\bibitem{zibold2010classical}
Zibold T, Nicklas E, Gross C and Oberthaler M~K 2010 Classical bifurcation at the transition from {R}abi to {J}osephson dynamics {\em Physical Review Letters\/} {\bf 105} 204101

\bibitem{schlagheck2022resurgent}
Schlagheck P, Ullmo D, Lando G~M and Tomsovic S 2022 Resurgent revivals in bosonic quantum gases: {A} striking signature of many-body quantum interferences {\em Physical Review A\/} {\bf 106} L051302

\bibitem{kalosakas2003multiple}
Kalosakas G, Bishop A and Kenkre V 2003 Multiple-timescale quantum dynamics of many interacting bosons in a dimer {\em Journal of Physics B: Atomic, Molecular and Optical Physics\/} {\bf 36} 3233--3238

\bibitem{pudlik2014tunneling}
Pudlik T, Hennig H, Witthaut D and Campbell D~K 2014 Tunneling in the self-trapped regime of a two-well {B}ose-{E}instein condensate {\em Physical Review A\/} {\bf 90} 053610

\bibitem{radcliffe1971some}
Radcliffe J~M 1971 Some properties of coherent spin states {\em Journal of Physics A\/} {\bf 4} 313

\bibitem{perelomov1977generalized}
Perelomov A~M 1977 Generalized coherent states and some of their applications {\em Soviet Physics Uspekhi\/} {\bf 20} 703

\bibitem{zhang1990coherent}
Zhang W~M, Gilmore R {\em et~al\/} 1990 Coherent states: {T}heory and some applications {\em Reviews of Modern Physics\/} {\bf 62} 867

\bibitem{buonsante2008some}
Buonsante P and Penna V 2008 Some remarks on the coherent-state variational approach to nonlinear boson models {\em Journal of Physics A\/} {\bf 41} 175301

\bibitem{trimborn2008exact}
Trimborn F, Witthaut D and Korsch H 2008 Exact number-conserving phase-space dynamics of the {M}-site {B}ose-{H}ubbard model {\em Physical Review A\/} {\bf 77} 043631

\bibitem{trimborn2009beyond}
Trimborn F, Witthaut D and Korsch H 2009 Beyond mean-field dynamics of small {B}ose-{H}ubbard systems based on the number-conserving phase-space approach {\em Physical Review A\/} {\bf 79} 013608

\bibitem{qiao2024quench}
Qiao Y and Grossmann F 2024 Quench dynamics of interacting bosons: generalized coherent states versus multi-mode {G}lauber states {\em Journal of Physics A\/} {\bf 57} 295302

\bibitem{vardi2001bose}
Vardi A and Anglin J 2001 Bose-{E}instein condensates beyond mean field theory: {Q}uantum backreaction as decoherence {\em Physical Review Letters\/} {\bf 86} 568

\bibitem{esteve2008squeezing}
Est{\`e}ve J, Gross C, Weller A, Giovanazzi S and Oberthaler M~K 2008 Squeezing and entanglement in a {B}ose-{E}instein condensate {\em Nature\/} {\bf 455} 1216

\bibitem{muessel2014scalable}
Muessel W, Strobel H, Linnemann D, Hume D~B and Oberthaler M~K 2014 Scalable spin squeezing for quantum-enhanced magnetometry with {B}ose-{E}instein condensates {\em Physical Review letters\/} {\bf 113} 103004

\bibitem{steel1998dynamical}
Steel M~J, Olsen M, Plimak L, Drummond P, Tan S, Collett M, Walls D and Graham R 1998 Dynamical quantum noise in trapped {B}ose-{E}instein condensates {\em Physical Review A\/} {\bf 58} 4824

\bibitem{sinatra2002truncated}
Sinatra A, Lobo C and Castin Y 2002 The truncated {W}igner method for {B}ose-condensed gases: limits of validity and applications {\em Journal of Physics B\/} {\bf 35} 3599

\bibitem{polkovnikov2010phase}
Polkovnikov A 2010 Phase space representation of quantum dynamics {\em Annals of Physics\/} {\bf 325} 1790

\bibitem{ray2016dynamics}
Ray S, Ostmann P, Simon L, Grossmann F and Strunz W~T 2016 Dynamics of interacting bosons using the {Herman-Kluk} semiclassical initial value representation {\em Journal of Physics A\/} {\bf 49} 165303

\bibitem{simon2014time}
Simon L and Strunz W~T 2014 Time-dependent semiclassics for ultracold bosons {\em Physical Review A\/} {\bf 89} 052112

\bibitem{engl2016semiclassical}
Engl T, Urbina J~D and Richter K 2016 The semiclassical propagator in {F}ock space: dynamical echo and many-body interference {\em Philosophical transactions A\/} {\bf 374} 20150159

\bibitem{mathew2019semiclassical}
Mathew R and Tiesinga E 2019 A semiclassical theory of phase-space dynamics of interacting bosons {\em Journal of Physics B\/} {\bf 52} 185302

\bibitem{wimberger2021finite}
Wimberger S, Manganelli G, Brollo A and Salasnich L 2021 Finite-size effects in a bosonic {J}osephson junction {\em Physical Review A\/} {\bf 103} 023326

\bibitem{shalashilin2001description}
Shalashilin D~V and Child M~S 2001 Description of tunneling with the help of coupled frozen {G}aussians {\em The Journal of Chemical Physics\/} {\bf 114} 9296

\bibitem{shalashilin2008gaussian}
Shalashilin D~V and Burghardt I 2008 Gaussian-based techniques for quantum propagation from the time-dependent variational principle: {F}ormulation in terms of trajectories of coupled classical and quantum variables {\em The Journal of Chemical Physics\/} {\bf 129}

\bibitem{qiao2021exact}
Qiao Y and Grossmann F 2021 Exact variational dynamics of the multimode {B}ose-{H}ubbard model based on {SU(M)} coherent states {\em Physical Review A\/} {\bf 103} 042209

\bibitem{qiao2023revealing}
Qiao Y and Grossmann F 2023 Revealing quantum effects in bosonic {J}osephson junctions: a multi-configuration atomic coherent state approach {\em Frontiers in Physics\/} {\bf 11} 1221614

\bibitem{heller1981frozen}
Heller E~J 1981 Frozen {G}aussians: {A} very simple semiclassical approximation {\em Journal of Chemical Physics\/} {\bf 75} 2923

\bibitem{littlejohn1986semiclassical}
Littlejohn R~G 1986 The semiclassical evolution of wave packets {\em Physics Reports\/} {\bf 138} 193

\bibitem{viscondi2011initial}
Viscondi T~F and de~Aguiar M~A 2011 Initial value representation for the $su(n)$ semiclassical propagator {\em The Journal of Chemical Physics\/} {\bf 134}

\bibitem{huber1988hybrid}
Huber D and Heller E~J 1988 Hybrid mechanics: {A} combination of classical and quantum mechanics {\em The Journal of Chemical Physics\/} {\bf 89} 4752

\bibitem{bergold2022error}
Bergold P and Lasser C 2022 An error bound for the time-sliced thawed {G}aussian propagation method {\em Numerische Mathematik\/} {\bf 152} 511

\bibitem{heller2018semiclassical}
Heller E~J 2018 {\em The semiclassical way to dynamics and spectroscopy\/} (Princeton University Press)

\bibitem{tatchen2011renormalization}
Tatchen J, Pollak E, Tao G and Miller W~H 2011 Renormalization of the frozen gaussian approximation to the quantum propagator {\em The Journal of chemical physics\/} {\bf 134}

\bibitem{abramowitz1948handbook}
Abramowitz M and Stegun I~A 1972 {\em Handbook of mathematical functions with formulas, graphs, and mathematical tables\/} (Dover Books on Advanced mathematics)

\bibitem{davis1984comparisons}
Davis M~J and Heller E 1984 Comparisons of classical and quantum dynamics for initially localized states {\em The Journal of chemical physics\/} {\bf 80} 5036

\bibitem{grossmann2000semiclassical}
Grossmann F 2000 Semiclassical real-time tunneling by multiple spawning of classical trajectories {\em Physical review letters\/} {\bf 85} 903

\bibitem{kroninger2023sampling}
Kr{\"o}ninger F, Lasser C and Van{\'\i}{\v{c}}ek J~J 2023 Sampling strategies for the {H}erman-{K}luk propagator of the wavefunction {\em Frontiers in Physics\/} {\bf 11} 1106324

\bibitem{graefe2008mean}
Graefe E~M, Korsch H and Niederle A 2008 Mean-field dynamics of a {non-Hermitian} {Bose-Hubbard} dimer {\em Physical Review Letters\/} {\bf 101} 150408

\bibitem{graefe2010quantum}
Graefe E~M, Korsch H~J and Niederle A~E 2010 Quantum-classical correspondence for a non-hermitian bose-hubbard dimer {\em Physical Review A\/} {\bf 82} 013629

\bibitem{graefe2013mean}
Graefe E~M and Liverani C 2013 Mean-field approximation for a bose--hubbard dimer with complex interaction strength {\em Journal of Physics A\/} {\bf 46} 455201

\bibitem{kramer2008review}
Kramer P 2008 A review of the time-dependent variational principle {\em Journal of Physics: Conference Series\/} {\bf 99} 012009

\bibitem{werther2021coherent}
Werther M, Choudhury S~L and Gro{\ss}mann F 2021 Coherent state based solutions of the time-dependent {S}chr{\"o}dinger equation: hierarchy of approximations to the variational principle {\em International Reviews in Physical Chemistry\/} {\bf 40} 81

\bibitem{bender1999advanced}
Bender C~M and Orszag S~A 1999 {\em Advanced mathematical methods for scientists and engineers: {A}symptotic methods and perturbation theory\/} vol~1 (Springer)

\bibitem{klimov2017generalized}
Klimov A~B, Romero J~L and De~Guise H 2017 Generalized {SU(2)} covariant {W}igner functions and some of their applications {\em Journal of Physics A\/} {\bf 50} 323001

\bibitem{edmonds1996angular}
Edmonds A~R 1996 {\em Angular momentum in quantum mechanics\/} vol~4 (Princeton university press)

\end{thebibliography}
\end{document}